\documentclass[prd,twocolumn,superscriptaddress,amsmath,amssymb,floatfix,nofootinbib]{revtex4-2}
\usepackage{graphicx}
\usepackage{booktabs}
\usepackage{xcolor}
\usepackage[colorlinks=true,allcolors=blue]{hyperref}
\newcommand{\um}{\,\mu\mathrm{m}}

\newcommand{\Halpha}{\mathrm{H}\alpha}
\newcommand{\Hbeta}{\mathrm{H}\beta}
\newcommand{\OIII}{[\textsc{O\,iii}]}
\newcommand{\OII}{[\textsc{O\,ii}]}
\newcommand{\Lya}{\mathrm{Ly}\alpha}
\newcommand{\nW}{\,\mathrm{nW\,m^{-2}\,sr^{-1}}}

\newcommand{\lamobs}{\lambda_{\mathrm{obs}}}
\newcommand{\Mpcinv}{\,\mathrm{Mpc}^{-1}}
\newsavebox{\fitbox}
\newcommand{\fitwidth}[1]{\sbox{\fitbox}{#1}%
  \ifdim\wd\fitbox>\linewidth\resizebox{\linewidth}{!}{\usebox{\fitbox}}\else\usebox{\fitbox}\fi}
\newcommand{\pIsci}[1]{\expandafter\pIsciA#1\relax}
\def\pIsciA#1e#2#3\relax{$#1\times10^{#2\number#3}$}

\newcommand{\XKIHLapow}{0.70}

\newcommand{\XKtwohalo}{83}

\newcommand{\XKtwohaloell}{350}

\newcommand{\XKtwohaloautoell}{43}

\newcommand{\XKnclusterlo}{1.5}
\newcommand{\XKnclustermid}{9.9}
\newcommand{\XKihlmedmass}{2.0\times10^{14}}

\newcommand{\XKcrossratlo}{1.5}
\newcommand{\XKcrossrathi}{4.2}
\newcommand{\XKautofactor}{1.5}

\newcommand{\XKzcrossdrop}{15}
\newcommand{\XKzautofac}{4.4}

\newcommand{\XKzIHLautomed}{0.03}
\newcommand{\XKzIHLautofa}{77.3}

\newcommand{\XKzIHLcrossfa}{14.7}

\newcommand{\XKzIHLcrossonehalomed}{0.07}

\newcommand{\XKzIHLcrosstwohalomed}{0.40}

\newcommand{\XKzIGLautomed}{0.63}

\newcommand{\XKzkkmed}{1.86}

\newcommand{\XKzIHLcrossonehalomedk}{0.23}

\newcommand{\XKzIHLcrosstwohalomedk}{0.62}

\newcommand{\RSihlellmin}{325}

\newcommand{\LKlam}{2.2}

\newcommand{\LKlyalamhi}{0.85}

\newcommand{\LKzmax}{6}

\newcommand{\LKtotrlo}{0.10}
\newcommand{\LKtotrhi}{0.16}

\newcommand{\LKlinepc}{0.47}

\newcommand{\EFoneIHLat}{1.2}

\newcommand{\EFoneIGLat}{3.0}

\newcommand{\EFtwoIHLavg}{5.2}

\newcommand{\EFtwoIGLavg}{3.3}

\newcommand{\EFtwoIHLcaoAvg}{0.66}

\newcommand{\GFivehikhi}{+43}

\newcommand{\GThreeZloMax}{+19.0}

\newcommand{\GThreeZfiveMin}{-24.6}

\newcommand{\CCfiveContMax}{8.1}

\newcommand{\CCsixContMedFour}{+9.5}

\newcommand{\CCsixContMedFive}{+7.9}

\begin{document}

\title{Cross-correlating the SPHEREx extragalactic background light with CMB lensing I:\\ theoretical predictions and forecasts}
\newcommand{\LAPTh}{Laboratoire d'Annecy de Physique Th\'eorique (LAPTh), CNRS/USMB, 99 Chemin de Bellevue BP110, Annecy F-74941, France}
\newcommand{\Sussex}{Department of Physics \& Astronomy, University of Sussex, Brighton BN1 9QH, UK}
\author{Azadeh Moradinezhad Dizgah}
\newcommand{\Montpolier}{Laboratoire Univers et Particules de Montpellier (LUPM), CC 72, Place Eug\`ene Bataillon, 34095 Montpellier - Cedex 5, France}
\newcommand{\UGA}{Univ. Grenoble Alpes, CNRS, Grenoble INP, LPSC-IN2P3, 53, avenue des Martyrs, 38000 Grenoble, France}
\email{azadeh.moradinezhad@lapth.cnrs.fr}
\affiliation{\LAPTh}
\author{Zucheng Gao}
\email{Zucheng.Gao@sussex.ac.uk}
\affiliation{\LAPTh}
\affiliation{\Sussex}
\author{Tom Chiesura \vspace{0.05in}}
\email{tom.chiesura@univ-grenoble-alpes.fr}
\affiliation{\LAPTh}
\affiliation{\Montpolier}
\affiliation{\UGA \vspace{0.1in}}

\begin{abstract}
The near-infrared extragalactic background light (EBL), which SPHEREx maps in 102 spectral channels, holds the emission
lines of galaxies on top of a much brighter continuum from the integrated galaxy light (IGL) and the intrahalo light
(IHL). Cross-correlating these maps with the lensing of the cosmic microwave background (CMB) isolates the emission that
traces the matter, unbiased by uncorrelated foregrounds. The lines face two obstacles. Filtering along
the line of sight removes the long radial modes that carry the cross-correlation, and the extragalactic continuum, a
signal in its own right, is a correlated foreground for the lines. We forecast the detectability of all EBL components, in a
common halo-model framework, for the SPHEREx deep fields and all-sky survey with lensing maps at the sensitivity level
of Planck and the Simons Observatory (SO). Without filtering, and with all amplitudes fitted jointly, the cross-correlation
with Planck lensing detects the IGL at $9\sigma$ in the deep fields, but not the lines. In the all-sky survey, it
measures the IGL and IHL amplitudes to $1.0\%$ and $3\%$ and detects the five lines, fitted as one template, at
$4.6\sigma$. With SO lensing, these become $0.6\%$, $1.6\%$ and $7.4\sigma$. In these joint fits, a filter removing radial modes longer than about $600$\,Mpc costs every component $90\%$ or more of its signal-to-noise ratio, so the lines are reached by
fitting the continuum, not by filtering it. These values assume a known spectral shape of the continuum. Calibrating it
with the continuum's auto-spectra keeps the lines at $7.3$--$7.4\sigma$; fitting it from the cross-spectrum alone leaves
$1.4$--$2.3\sigma$. The line detections vary by a factor of seven to eleven across line models. The cross-correlation is
thus a robust probe of the IGL and a viable route to the emission lines, provided the continuum is
modeled across the channels.
\end{abstract}

\maketitle

\section{Introduction}
\label{sec:intro}

The extragalactic background light (EBL) is the summed emission of all sources over cosmic history. In the near
infrared, its continuum has two parts. The integrated galaxy light (IGL) gathers the starlight of every galaxy,
detected or not, and so measures the total output of cosmic star formation; galaxy counts only bound it from below.
The intrahalo light (IHL) comes from stars stripped from their host galaxies during mergers and tidal
interactions~\cite{Purcell2007,Cooray2012}, and is starlight that no galaxy census contains. On this continuum, star-forming
galaxies imprint their emission lines, a small share of the light, which measure the star formation at the redshift set by the observed wavelength. Because all three trace the large-scale distribution of the dark matter, they are not
only astrophysical but also, in principle, cosmological probes, the continuum as a cumulative signal along the line of sight (LoS) and the lines as a tomography in redshift. Both uses, however, require the components to be separated, and the fluctuations in a single band do not reveal their source. Once the resolved galaxies are masked, the measured
fluctuation power exceeds what the faint galaxies that remain can
supply~\cite{Kashlinsky2005,Kashlinsky2012,Matsumoto2011,MitchellWynne2015,Zemcov2014,Kashlinsky2018,Bock2025}.
Whether the excess comes from the IHL~\cite{Cooray2012,Zemcov2014,Cheng2021} or from high-redshift
sources~\cite{Kashlinsky2018} remains debated (see Ref.~\cite{Cooray2016} for a review). The separation of the components relies on
several bands and on cross-correlations with other
tracers~\cite{Feng2019,EuclidCollab2026}, among them galaxies of known redshift, which slice the background in
redshift~\cite{Chiang2019,ChengChang2021,Scott2022}.

NASA's SPHEREx satellite~\cite{Dore2014,Dore2018,Crill2020} is now mapping the whole sky in 102 spectral channels
between $0.75$ and $5\um$, and observes two deep fields of $100\,{\rm deg}^2$ each near the ecliptic
poles~\cite{Bock2025,Korngut2026}. The deep fields are observed fifty to a hundred times more often than the rest of the
sky~\cite{Bock2025}. Each channel records all the extragalactic light at its wavelength, including the emission lines
of galaxies at matching redshifts. These lines, H$\alpha$, H$\beta$, \OIII, \OII{} and
Ly$\alpha$, are the targets of line-intensity mapping, which measures the aggregate emission of all galaxies in a
line rather than resolving individual sources (see Refs.~\cite{Kovetz2017,Bernal2022,ChangLidz2026} for reviews).
Forecasts for SPHEREx and similar surveys have shown that the clustering of these lines traces star formation and
large-scale structure over a wide range of
redshifts~\cite{Pullen2014,Fonseca2017,Gong2017,Silva2018,Gong2020,Heneka2021,Cheng2024,Ambrose2026,Roy2026b}.
In every channel, however, the lines are superposed on the far brighter continuum of the IGL and IHL.\looseness=-1

Cross-correlating intensity maps with the convergence $\kappa$ of the gravitational lensing of the cosmic microwave
background (CMB)~\cite{LewisChallinor2006} isolates the part of the signal that traces the matter
distribution. CMB-lensing maps have improved steadily since the Planck measurement~\cite{Planck2018lensing}, with the
Atacama Cosmology Telescope (ACT)~\cite{Madhavacheril2024,Qu2024} and the South Pole Telescope (SPT)~\cite{Ge2025}, and are
expected to improve considerably further with the Simons Observatory (SO)~\cite{Ade2019}. Their cross-correlation with
the cosmic infrared background (CIB), the thermal emission of dust heated by young stars in star-forming galaxies, was
predicted~\cite{Song2003} and then detected at high significance~\cite{Planck2014CIB,Holder2013,vanEngelen2015,Cao2020}.
It has since been used to constrain models of the emitting galaxies~\cite{McCarthy2021}. In the near infrared, local foregrounds and instrumental systematics would enter
the variance of the cross-correlation with lensing, but not its mean. Reference~\cite{Feder2025} anticipated it as a test of the
nature of the excess fluctuations.

For line-intensity maps, lensing has mostly been studied as a signal in the maps themselves: the lensing convergence
can be reconstructed from them~\cite{Pourtsidou2014,Foreman2018,Schaan2018}, and cross-correlated with CMB lensing to
null the contribution of low redshifts~\cite{Maniyar2022,Fronenberg2024a,Fronenberg2024b}. The cross-correlation of the
line intensity itself with lensing has been forecast for CO maps with cosmic shear, as a probe of the cosmic molecular
gas density~\cite{Chung2022}. For the SPHEREx lines, it has been forecast in searches for decaying axion-like
particles, with galaxy lensing~\cite{Shirasaki2021} and with galaxy clustering, galaxy lensing and CMB
lensing~\cite{WuXia2025}, where the emission lines are detected beside the axion signal.

The direct cross-correlation of a line-intensity map with CMB lensing faces an obstacle,
however~\cite{SchaanWhite2021b}. Spectrally smooth foregrounds are removed by discarding the slowly varying modes along
each LoS, either from the maps themselves or from the measured three-dimensional power spectrum. The
lensing kernel, smooth over gigaparsecs, carries only modes whose wavenumber along the LoS, $k_\parallel$, is close to zero. Several works therefore
concluded that this cross-correlation is beyond reach in practice~\cite{Obuljen2018,Li2019,Modi2019LSS,Guandalin2022,Moodley2023},
and proposed to recover the lost modes through nonlinear reconstruction~\cite{Zhu2018,Li2019,Modi2019LSS} or
higher-order statistics~\cite{Moodley2023,CHIME2026}. Numerical studies have since revisited the
question~\cite{Sangka2024,Marins2025}. Recently, Ref.~\cite{Shen2025} showed that the growth of structure and the evolution of the
line emission along the lightcone couple modes along the LoS and keep part of the filtered cross-correlation. They
forecast its detectability for several spectral lines, including Ly$\alpha$ with SPHEREx, and concluded that the
cross-correlation is within reach of future wide-field line surveys but probably not of current experiments.

The forecasts of the SPHEREx lines' cross-correlation with lensing treat the lines alone, without the continuum.
Those of Refs.~\cite{Shirasaki2021,WuXia2025} apply no filter along the LoS, and Ref.~\cite{Shen2025} sets the
spectrally smooth extragalactic background aside as a different kind of foreground. For the line maps themselves, the
continuum has been modeled together with the lines and removed by principal components~\cite{Roy2026a} or by neural
networks~\cite{Cagliari2025}. In the cross-correlation with lensing, however, the continuum of a SPHEREx channel cannot
be set aside: it is extragalactic, correlates with $\kappa$, and remains in the mean of the measured cross-spectrum.

In this paper, we compute the cross-spectra of every component of a SPHEREx channel with CMB lensing and forecast
their detectability, treating the lines and the continuum together. We model the IGL, the IHL and the five emission
lines in a common halo model, and compute their auto- and cross-spectra on the 102 SPHEREx channels. We apply to these
channels the filter along the LoS that an intensity-mapping analysis would use, and follow what it does to
each component, in the auto-spectra and in the cross-spectra with lensing. We then forecast the detectability with the
statistic of Ref.~\cite{Shen2025}, for the deep fields and the all-sky survey at the in-flight sensitivity of
SPHEREx~\cite{Bock2025}. The lensing maps have the noise of Planck, of SO or of a hypothetical map with a fifth of the
SO noise. Instead of filtering the continuum out, we fit the amplitudes of all components jointly. We also ask how well
the spectral shape of the continuum must be known for the lines to be detected: known in advance, calibrated by the
continuum's own auto-spectra, or fitted from the cross-spectrum alone. We evaluate the line auto-spectra in the flat-sky projection of
Ref.~\cite{Gao2023}, and verify that the Limber approximation is accurate for every cross-spectrum with $\kappa$. We
limit our analysis to two-point statistics with a Gaussian covariance. In Paper~II of this series~\cite{PaperII}, we build the same
observables from simulated maps of SPHEREx and CMB lensing.

The rest of the paper is organized as follows. In Sec.~\ref{sec:model}, we describe the model of the angular spectra,
the filter along the LoS and the halo model of the components, and in Sec.~\ref{sec:surveys} the survey
configurations and the statistics of the forecast. We present the unfiltered cross-spectra in Sec.~\ref{sec:unfiltered}
and the effect of the filter in Sec.~\ref{sec:filtered}, forecast the detectability in Sec.~\ref{sec:snr}, and
conclude in Sec.~\ref{sec:conclusions}. The appendices validate the model against the literature
(Appendix~\ref{app:validation}) and show which component each lensing map selects (Appendix~\ref{app:shear}). They
detail the flat-sky and Limber evaluations and the filter conventions (Appendix~\ref{app:flatsky}), compare with
Ref.~\cite{Shen2025} (Appendix~\ref{app:shen}), and list the detectability of every component
(Appendix~\ref{app:surveytable}).

Throughout, we assume a flat $\Lambda$CDM cosmology with $\Omega_m = 0.309$, $\Omega_b = 0.049$, $h = 0.677$,
$\sigma_8 = 0.822$ and $n_s = 0.968$, that of the Hidden Valley simulations~\cite{Modi2019} from which Paper~II
builds its maps.
\section{Modeling the cross-correlation}
\label{sec:model}

In this section, we describe how we model the angular cross-spectrum between a SPHEREx channel map and the CMB-lensing
convergence.

\subsection{Angular power spectra}
\label{sec:projection}

Our observables are the angular power spectra of the EBL components in each SPHEREx channel, and their
cross-spectra with the CMB-lensing convergence. Each is the spectrum of two projected fields. A projected field is the
integral along the LoS of a three-dimensional fluctuation $\delta_X$, weighted by a radial kernel $W_X$,
\begin{equation}
  X(\hat{\mathbf n}) = \int\! d\chi\, W_X(\chi)\, \delta_X(\chi\hat{\mathbf n};z),
  \label{eq:projected}
\end{equation}
where $z$ is the redshift at the comoving distance $\chi$. With $P_{XY}(k;z,z')$ the unequal-time cross-power spectrum
of $\delta_X$ and $\delta_Y$, the angular cross-spectrum on the full sky is
\begin{equation}
  \begin{split}
  C_\ell^{XY} ={}& \frac{2}{\pi}\int_0^\infty\! dk\,k^2 \int\! d\chi\,W_X(\chi)\int\! d\chi'\,W_Y(\chi')\\
  &\times j_\ell(k\chi)\,j_\ell(k\chi')\,P_{XY}(k;z,z'),
  \end{split}
  \label{eq:fullsky}
\end{equation}
where $j_\ell$ are the spherical Bessel functions. For the convergence, $\delta_Y$ is the matter density contrast and
the kernel is $W_\kappa(\chi) = \tfrac32\Omega_m (H_0/c)^2 (1+z)\,\chi\,(\chi_\star-\chi)/\chi_\star$, with $\chi_\star$
the comoving distance to the last-scattering surface. The spectrum $P_{XY}$ is then the cross-power spectrum
$P_{X\delta}$ of component $X$ with matter. The kernels and power spectra of the components are given in Sec.~\ref{sec:components}.

Equation~\eqref{eq:fullsky} is rarely evaluated directly, because its integrals over the highly oscillating Bessel
functions are numerically costly, and most analyses adopt the Limber approximation instead. When both kernels vary
slowly over the correlation length $\chi/\ell$, the Bessel functions confine the pair to $\chi \approx \chi'$ and the
wavenumber to $k = (\ell+\tfrac12)/\chi$. The cross-spectrum with $\kappa$ then reduces to
\begin{equation}
  C_\ell^{\kappa X} = \int\! d\chi\,
    \frac{W_X(\chi)\,W_\kappa(\chi)}{\chi^2}\,
    P_{X\delta}\!\left(k = \frac{\ell+\tfrac12}{\chi},\, z\right),
  \label{eq:xkappa}
\end{equation}
with $W_X^2$ and $P_{XX}$ for an auto-spectrum.

The Limber approximation fails when a kernel varies over the correlation length, that is, for narrow kernels and at low
multipoles. The thin shell of an emission line in one channel is such a kernel, and Ref.~\cite{Shen2025} argued that a
filter along the LoS creates radial structure of the same kind. Since the Limber approximation sets the wavenumber
along the LoS to zero, it also misses the redshift-space distortions, which depend on that wavenumber. The flat-sky
approximation of Ref.~\cite{Gao2023} removes these limitations. It replaces the spherical harmonics by plane waves but retains the wavenumber along the LoS,
$k_\parallel$, and the two epochs of the pair. In the tests of Ref.~\cite{Gao2023}, it agrees with the full-sky result
at the percent level or better. With the pair placed at $\chi = \bar\chi + \delta\chi/2$ and $\chi' = \bar\chi - \delta\chi/2$, the flat-sky spectrum reads
\begin{equation}
  \begin{split}
  C_\ell^{XY} ={}& \int\! \frac{d\bar\chi}{\bar\chi^2}\int\! d\delta\chi\; W_X(\chi)\,W_Y(\chi')\\
  &\times\int_{-\infty}^{\infty}\!\frac{dk_\parallel}{2\pi}\,e^{ik_\parallel\delta\chi}\\
  &\times P_{XY}\!\left(\sqrt{k_\parallel^2 + \tilde\ell^{\,2}};\, z, z'\right),
  \end{split}
  \label{eq:flatsky}
\end{equation}
where $\tilde\ell^{\,2} = \ell(\ell+1)/[\bar\chi^2(1-x^2)]$, with $x = \delta\chi/(2\bar\chi)$, is the geometric
recalibration of the multipole introduced in Ref.~\cite{Gao2023}. Dropping the factor $(1-x^2)$ changes
our results by less than $10^{-4}$. When both kernels vary slowly over $\chi/\ell$, the $k_\parallel$ integral confines
the pair to $\delta\chi \approx 0$, and Eq.~\eqref{eq:flatsky} reduces to the Limber approximation.

The flat-sky integral can be computed efficiently, as in Ref.~\cite{Gao2023}, with the FFTLog algorithm, which writes the power spectrum as a sum of power laws whose $k_\parallel$ integrals are known in closed form. In our
implementation, we repeat the decomposition at each mean distance, since a halo-model spectrum is not separable in
scale and time, and we carry the unequal-time dependence of the two-halo term as a scale-independent ratio of growth factors and biases. For
the emission lines we include linear redshift-space distortions~\cite{Kaiser1987}. The details of the evaluation and
its validation are given in Appendix~\ref{app:flatsky}.

Comparing the two evaluations, we find that every cross-spectrum with $\kappa$ agrees with its Limber value to $0.3\%$
or better. The lensing kernel varies slowly over $\chi/\ell$ at every multipole of interest, and it confines the pair to
$\delta\chi \approx 0$ even when the other kernel is narrow. The auto-spectrum of an emission line in one channel is the
single exception: its kernel is a thin shell with sharp edges, and the Limber approximation overestimates the spectrum by $42\%$
at $\ell \approx 90$ (Appendix~\ref{app:limber}). We therefore use the Limber
approximation for the continuum and the flat-sky projection of Eq.~\eqref{eq:flatsky} for the
lines.\footnote{Reference~\cite{Shen2025} found that the Limber approximation fails for filtered
cross-spectra. We return to this in Sec.~\ref{sec:filtered} and Appendix~\ref{app:shen}.}

Throughout the paper, we quote the spectra as bandpowers,
\begin{equation}
  D_\ell \equiv \frac{\ell(\ell+1)}{2\pi}\,C_\ell ,
  \label{eq:Dell}
\end{equation}
in units of $(\mathrm{nW\,m^{-2}\,sr^{-1}})^2$ for an intensity auto-spectrum and of $\nW$ for a cross-spectrum with the dimensionless convergence $\kappa$.

\subsection{Filtering along the LoS}
\label{sec:filter}

Spectrally smooth foregrounds are removed from intensity maps by discarding the slowly varying modes along each LoS. A
SPHEREx channel, however, is a wavelength, not a distance: each emission line places it at a different redshift, and
the continuum in a channel integrates over all distances. Modes along the LoS therefore require a radial coordinate,
and we use the comoving distance $\chi_\nu$ at which H$\alpha$ falls in channel $\nu$. H$\alpha$ is the brightest line
and spans the whole band. For H$\alpha$, the 102 channels are consecutive shells in distance, which together cover
$z = 0.14$--$6.6$, a depth of $8.1$\,Gpc. The filter removes the Fourier modes over this depth with wavenumbers below a
chosen filter scale. For the other lines these modes are smooth functions of their own distance, and for the continuum
smooth functions of wavelength. With $\chi_{\min}$ and $\chi_{\max}$ the distances of the bluest and reddest channels and $\Delta\chi_\nu = \chi_\nu - \chi_{\min}$, the fundamental wavenumber of the band is $k_F =
2\pi/(\chi_{\max} - \chi_{\min}) = 7.8\times10^{-4}\Mpcinv$.

A high-pass filter at scale $\Lambda$ is then the projection
\begin{equation}
  \begin{aligned}
  \mathsf{H} &= \mathbb{I} - QQ^{\sf T},\qquad m = \lfloor \Lambda/k_F \rfloor,\\
  Q &= {\rm orth}\bigl\{1,\ \cos(jk_F\Delta\chi_\nu),\ \sin(jk_F\Delta\chi_\nu)\bigr\}_{j=1}^{m},
  \end{aligned}
  \label{eq:projection}
\end{equation}
which removes the constant and the $m$ lowest cosine and sine modes of the band, $2m+1$ modes in total. Evaluated at
the channel distances, each mode is a vector over the 102 channels. Since the channels are unevenly spaced in $\chi$,
these vectors are not orthogonal, and ${\rm orth}$ denotes their orthonormalization, by a QR decomposition, into the
$2m+1$ columns of the $102\times(2m+1)$ matrix $Q$. The filter $\mathsf{H}$ is then a $102\times102$ matrix, with
elements $\mathsf{H}_{\nu\nu'}$, that subtracts from the 102 channel values of every LoS their least-squares fit by the
$2m+1$ modes. At $\Lambda = 0.01\Mpcinv$, $m = 12$ and the filter removes $25$ of the $102$ channel directions, the independent linear combinations of the channel values; at $m =
0$ it removes only the mean over the channels, and from $\Lambda \approx 0.04\Mpcinv$, where $2m+1$ exceeds the
number of channels, it removes every direction. This limit reflects how the channels sample the LoS. Adjacent channels lie
about $100$\,Mpc apart over most of the band, so that by $\Lambda \approx 0.03\Mpcinv$ the filter has removed nearly
every direction these channels can represent. Only the channels at the red end, $22$--$28$\,Mpc apart, keep
directions up to $0.04\Mpcinv$, the Nyquist wavenumber of the mean channel spacing of $80$\,Mpc.

The high-pass filter of Ref.~\cite{Shen2025} removes the same modes, with the same $m$ (their Eqs.~(25)--(30)). They apply it to the continuous intensity along the LoS, before binning it in distance, whereas the
projection acts on the channel values themselves and removes exactly $2m+1$ degrees of freedom of the data. The two
coincide for many, evenly spaced bins and differ when the bins are few, because filtering before binning lets part of
the removed modes return into the bins. On the 15 bins of their SPHEREx Ly$\alpha$ forecast, the projection gives a
signal-to-noise ratio $4\%$ lower at $m = 2$ and $27\%$ lower at $m = 5$, and leaves no data from $m = 7$. These bins, each about one channel deep ($72$\,Mpc), thus meet the same limit as the SPHEREx channels, since $m = 7$ corresponds to $\Lambda \approx 0.04\Mpcinv$ for their redshift range. With the filter of Ref.~\cite{Shen2025}, in contrast, the bins keep 14 independent values at every $\Lambda$ (their App.~D7), so that the SPHEREx forecast of that reference extends to $\Lambda = 0.1\Mpcinv$, the value they motivate by the foreground removal of 21-cm maps (their footnote~11). Finally, each line is filtered there over its own redshift range, whereas we filter every component over the whole band.

We apply the filter to the maps in specific intensity $I_\nu$, identically for every component and every statistic.
Continuum treatments proposed for SPHEREx remove smooth modes in other bases, either by fitting polynomials along the
spectral axis of each LoS~\cite{Gong2017,Cheng2020} or by removing the leading principal components of the maps (Gao
et al.~\cite{Gao2026}, as used in Ref.~\cite{Cagliari2025}). We compare these choices in Sec.~\ref{sec:channelmodes}.

Since the filter is linear in the maps, it acts on the spectra of Sec.~\ref{sec:projection} through the channel index.
With $W_X(\chi;\nu)$ the kernel of component $X$ in channel $\nu$, Eq.~\eqref{eq:flatsky} gives the cross-spectrum
profile $C_\ell^{\kappa X}(\nu)$, one value per channel, and the channel--channel covariance $C_\ell^{XX}(\nu,\nu')$.
With filtered quantities denoted by a tilde, the filter acts once on the former and on both indices of the latter,
\begin{equation}
  \begin{aligned}
  \tilde C_\ell^{\kappa X}(\nu) &= \sum_{\nu'} \mathsf{H}_{\nu\nu'}\,C_\ell^{\kappa X}(\nu'),\\
  \tilde C_\ell^{XX}(\nu,\nu') &= \sum_{\mu,\mu'} \mathsf{H}_{\nu\mu}\,C_\ell^{XX}(\mu,\mu')\,\mathsf{H}_{\nu'\mu'}.
  \end{aligned}
  \label{eq:filtobs}
\end{equation}
These relations filter the angular spectra after the integral along $\chi$ that produces them. The intensity-mapping
literature often filters the three-dimensional field before that integral instead. The filter is then idealized as a
convolution along an unbounded LoS (Ref.~\cite{Shen2025} uses the discrete Fourier transform of the kernel over the
finite window, their App.~C), with a response $H(k_\parallel) = \Theta(|k_\parallel| - \Lambda)$ that enters
the $k_\parallel$ integral of Eq.~\eqref{eq:flatsky} once for a cross-spectrum and squared for an auto-spectrum. For an emission line, whose channel kernel is a thin shell
in $\chi$, the channel index is the distance, and both forms filter the same three-dimensional field along the LoS. We
nevertheless use Eq.~\eqref{eq:filtobs}, for two reasons. First, for the continuum, whose every channel integrates the
emission over all distances, no such response exists: filtering inside the radial integral would filter a
three-dimensional field that is never observed. Second, a convolution along an unbounded LoS does not see the edges of
the band, which carry most of what survives the projection (Sec.~\ref{sec:filtered}). Alternative conventions for
the filter are compared in Appendix~\ref{app:filters}.\looseness=-1

\subsection{Halo model of the components}
\label{sec:components}

We describe the three components of the background in a common halo model~\cite{CooraySheth2002}, following
Refs.~\cite{Helgason2012,ChengChang2021} for the luminosities of the IGL, Ref.~\cite{EuclidCollab2026} for its
clustering and for the IHL, and Ref.~\cite{Gong2017} for the emission lines. We use the halo mass function and bias of
Ref.~\cite{ShethTormen1999} over the mass range $10^{8}$--$10^{16}\,M_\odot$, and Navarro--Frenk--White profiles~\cite{Navarro1997}
for halos of 200 times the mean matter density, with the concentration $c = 9\,(1+z)^{-1}(M/M_*)^{-0.13}$ of
Ref.~\cite{Bullock2001}, where $M_*$ is the nonlinear mass.

The cross-spectrum of each component with $\kappa$ is Eq.~\eqref{eq:xkappa}, with the kernel $W_X$ and the
cross-power spectrum with matter $P_{X\delta}$ given below. That cross-power spectrum follows from the component's
auto-spectrum by replacing one of its luminosity weights by the matter weight $M/\bar\rho_m$. In the one-halo term the
replacement acts inside the mass integral, and in the two-halo term it replaces one of the two mass integrals by the
matter bias integral
\begin{equation}
  \mathcal{I}_\delta(k,z) = \int\! dM\,n(M,z)\,\frac{M}{\bar\rho_m}\,b(M,z)\,u(k|M),
  \label{eq:Idelta}
\end{equation}
where $n(M,z) \equiv dn/dM$ is the halo mass function, $b(M,z)$ the halo bias and $u(k|M)$ the normalized Fourier transform of the
halo profile. For the two-halo
term to recover the linear matter power spectrum on large scales, $\mathcal{I}_\delta$ must tend to unity as $k \to
0$~\cite{CooraySheth2002}. We therefore assign to the mass below our lowest halo mass a unit bias and no internal
structure. Without this correction, the two-halo term of the IGL cross-spectrum would be $37$--$42\%$ too low.

\paragraph{Integrated galaxy light.}
The IGL is the integrated starlight of galaxies. Stellar photospheres emit nearly as blackbodies, so that this light is smooth in wavelength. In the rest-frame near infrared it is dominated by cool, old giant stars, so it traces the
stellar mass that a galaxy has built up over its history rather than its current star formation. Because the continuum is smooth, every galaxy along the LoS
contributes to every channel, at the rest-frame wavelength $\lamobs/(1+z)$, and the IGL kernel spans the whole
lightcone. Galaxies bright enough to be detected individually are masked in the maps, and the IGL of this paper is
the unresolved one, the light of the galaxies fainter than the masking threshold; the simulated maps of
Paper~II~\cite{PaperII} carry the total IGL instead, with no galaxy masked. The kernel of the unresolved IGL is the intensity $\nu I_\nu$ that the unresolved galaxies produce per
unit comoving distance,
\begin{equation}
  W_{\rm IGL}(z) = \chi^2\int_{m_{\rm lim}(z)}^{\infty}\! dm_{\rm AB}\;\nu f_\nu(m_{\rm AB})\,\Phi(M_{\rm abs}|z),
  \label{eq:Wigl}
\end{equation}
where $m_{\rm AB} = -2.5\log_{10}(f_\nu/3631\,{\rm Jy})$ is the apparent AB magnitude of a galaxy of flux density
$f_\nu$, so that $f_\nu(m_{\rm AB}) = 10^{-0.4(m_{\rm AB}-23.9)}\,\mu{\rm Jy}$. The absolute magnitude is $M_{\rm abs}
= m_{\rm AB} - {\rm DM}(z) + 2.5\log_{10}(1+z)$, with ${\rm DM}$ the distance modulus, and $\Phi$ is the Schechter
luminosity function of Ref.~\cite{Helgason2012}, evaluated at the rest-frame wavelength of the emission. Following
Ref.~\cite{ChengChang2021}, we define the masking once, by a catalog complete to $20$\,AB at $1\um$. In each channel,
$m_{\rm lim}(z)$ is the magnitude at which the number density of masked galaxies equals that of galaxies brighter than
$20$\,AB at $1\um$, so that every channel masks the same galaxies. The clustering of the IGL is described by the halo-occupation model of
Refs.~\cite{Zheng2005,Helgason2012},
\begin{align}
  P^{1h}_{\rm IGL}(k) &= \frac{1}{\bar n^2}\int\! dM\,n(M,z)\Bigl[2N_{\rm cen}N_{\rm sat}\,u(k|M)\nonumber\\
  &\qquad + N_{\rm sat}^2\,u^2(k|M)\Bigr],
  \label{eq:P1higl}\\
  P^{2h}_{\rm IGL}(k) &= P_{\rm lin}(k)\Bigl[\frac{1}{\bar n}\int\! dM\,n(M,z)\,N_{\rm gal}\nonumber\\
  &\qquad\qquad\times b(M,z)\,u(k|M)\Bigr]^2 ,
  \label{eq:P2higl}
\end{align}
where $N_{\rm gal} = N_{\rm cen} + N_{\rm sat}$, $\bar n = \int dM\,n(M,z)\,N_{\rm gal}$, and the mean numbers of
central and satellite galaxies in a halo of mass $M$ are~\cite{Zheng2005,Helgason2012,EuclidCollab2026}
\begin{align}
  N_{\rm cen}(M) &= \frac12\left[1 + {\rm erf}\!\left(\frac{\log_{10}(M/M_{\min})}{\sigma_{\log M}}\right)\right],
  \label{eq:Ncen}\\
  N_{\rm sat}(M) &= N_{\rm cen}(M/2)\left(\frac{M}{M_s}\right)^{\alpha_s},
  \label{eq:Nsat}
\end{align}
with $\sigma_{\log M} = 0.2$, $M_{\min} = 10^{9}\,M_\odot$, $M_s = 5\times10^{10}\,M_\odot$ and $\alpha_s = 1$, the values
adopted in Ref.~\cite{EuclidCollab2026}. The satellites switch on at twice the central threshold mass and grow as
a power law in mass. The cross-power spectrum with matter is~\cite{CooraySheth2002}
\begin{align}
  P^{1h}_{{\rm IGL},\delta}(k) &= \frac{1}{\bar n}\int\! dM\,n(M,z)\,\frac{M}{\bar\rho_m}\,u(k|M)\nonumber\\
  &\qquad\times\bigl[N_{\rm cen} + N_{\rm sat}\,u(k|M)\bigr],
  \label{eq:P1higlx}\\
  P^{2h}_{{\rm IGL},\delta}(k) &= P_{\rm lin}(k)\,\mathcal{I}_\delta(k)\nonumber\\
  &\qquad\times\frac{1}{\bar n}\int\! dM\,n(M,z)\,N_{\rm gal}\,b(M,z)\,u(k|M) ,
  \label{eq:P2higlx}
\end{align}
where in the one-halo term the central galaxy sits at the halo center and correlates with the matter profile of its own
halo through one power of $u(k|M)$, while the satellites follow the profile themselves. The auto-spectrum also contains the
Poisson term of the unresolved galaxies, the self-correlation of galaxies that are points at every SPHEREx scale,
\begin{equation}
  C_\ell^{\rm P} = \int\! dz\,\frac{d\chi}{dz}\,\chi^2 \int_{m_{\rm lim}(z)}^{\infty}\! dm_{\rm AB}\,
  \big[\nu f_\nu(m_{\rm AB})\big]^2\,\Phi(M_{\rm abs}|z) ,
  \label{eq:Pigl}
\end{equation}
constant in $\ell$, as in Refs.~\cite{ChengChang2021,EuclidCollab2026}. In the channel covariance of Sec.~\ref{sec:snrstat} this term has rank one, since the model assigns one spectrum to the unresolved
galaxies.

\paragraph{Intrahalo light.}
The IHL is the diffuse light of stars that are no longer bound to any galaxy. Satellite galaxies orbiting inside a larger halo
are stripped and eventually disrupted by its tidal field~\cite{Purcell2007}, and their stars then orbit in the
potential of the host halo, spread through it at very low surface brightness. These stars are mostly old, and no stars
form among them, so the IHL has a continuum but no emission lines. Being diffuse, it cannot be masked. Like that of the IGL, its kernel spans the whole lightcone. For the IHL
we adopt the model of Ref.~\cite{EuclidCollab2026},
\begin{align}
  W_{\rm IHL}(z) &= \frac{a^{2-\alpha}}{4\pi}\,S_{\rm IHL}\!\left(\frac{\lamobs}{1+z}\right),
  \label{eq:Wihl}\\
  P^{1h}_{\rm IHL}(k) &= \int\! dM\,n(M,z)\,f_{\rm IHL}^2\,L^2(M,0)\,u^2(k|M) ,
  \label{eq:P1hihl}\\
  P^{2h}_{\rm IHL}(k) &= P_{\rm lin}(k)\Bigl[\int\! dM\,n(M,z)\,f_{\rm IHL}\,L(M,0)\nonumber\\
  &\qquad\times b\,u(k|M)\Bigr]^2 ,
  \label{eq:P2hihl}
\end{align}
where $a = (1+z)^{-1}$ and $S_{\rm IHL}$ is the rest-frame spectrum, recovered by inverting the IHL kernels of Fig.~1 of
Ref.~\cite{EuclidCollab2026} through their Eq.~(5).
The one-halo term of the IHL is the self-correlation of one extended object per halo, so no Poisson term is added. Its cross-power
spectrum with matter is
\begin{align}
  P^{1h}_{{\rm IHL},\delta}(k) &= \int\! dM\,n(M,z)\,f_{\rm IHL}\,L(M,0)\,\frac{M}{\bar\rho_m}\,u^2(k|M) ,
  \label{eq:P1hihlx}\\
  P^{2h}_{{\rm IHL},\delta}(k) &= P_{\rm lin}(k)\,\mathcal{I}_\delta(k)\nonumber\\
  &\qquad\times\int\! dM\,n(M,z)\,f_{\rm IHL}\,L(M,0)\nonumber\\
  &\qquad\qquad\times b(M,z)\,u(k|M) ,
  \label{eq:P2hihlx}
\end{align}
with the light and the matter both on the halo profile.
The fraction of a halo's light in the diffuse component and the $K$-band luminosity of the halo are
\begin{align}
  f_{\rm IHL}(M) &= A_{\rm IHL}\left(\frac{M}{10^{12}\,M_\odot}\right)^{\beta},
  \label{eq:fihl}\\
  L(M,0) &= 2.76\times10^{12}\,h^{-2}L_{K,\odot}\nonumber\\
  &\quad\times\left(\frac{M}{1.9\times10^{14}\,h^{-1}M_\odot}\right)^{0.72},
  \label{eq:LM0}
\end{align}
where $A_{\rm IHL}$ and $\beta$ are fitted in seven mass bins, and we take the best fit of the joint analysis of the
EBL, cosmic shear and galaxy clustering in Ref.~\cite{EuclidCollab2026}. The luminosity is assumed to be separable,
$L(M,z) = L(M,0)(1+z)^{\alpha}$ with $\alpha = -1.05$, as in Refs.~\cite{Cooray2012,EuclidCollab2026}, so that its
redshift evolution enters through the kernel. Following Ref.~\cite{EuclidCollab2026}, the IHL integrals run over
$10^{10}$--$10^{15}\,M_\odot$.

Our continuum model differs from that of Ref.~\cite{EuclidCollab2026} in three choices, discussed in
Appendix~\ref{app:validation}. First, we evaluate the luminosity function in the rest frame and mask at $20$\,AB at $1\um$, whereas their Table~1 and Sec.~2.1.2 imply that they evaluate it in the observed band and mask at $20.5$\,AB in every band. The IGL in our convention is
$0.5$--$2.6$ times theirs across the band, and their convention masks a different galaxy population in every channel, which biases the lines (Sec.~\ref{sec:snrshape}). Second, our IHL kernel carries $a^{2-\alpha}$ instead of the $a^{1-\alpha}$ they print. Third, we take $L(M,0)$ in solar $K$-band units, whereas
their text calls $L(M,0)$ a total solar luminosity. With the first two of their conventions, our model does not reproduce the
amplitudes of the IGL and IHL auto-spectra in their Fig.~2 (Appendix~\ref{app:validation}).

\paragraph{Emission lines.}
The emission lines are produced by the ionized gas of star-forming regions. Only stars heavier than about ten solar masses, which live less than
about twenty million years, produce a significant number of the photons that ionize hydrogen~\cite{Kennicutt1998},
and they ionize the gas clouds in which they formed.
Recombining electrons cascade through the energy levels of hydrogen and emit the H$\alpha$, H$\beta$ and Ly$\alpha$
lines, and collisions excite oxygen ions, which emit the \OII{} and \OIII{} lines. The lines thus trace the current
star-formation rate, whereas the near-infrared continuum traces the stellar mass. A line falls in a given channel only
from galaxies in a thin shell, at $z = \lamobs/\lambda_{\rm line} - 1$ for a line of rest wavelength
$\lambda_{\rm line}$. For the emission lines we adopt the model of Ref.~\cite{Gong2017}, as in
Ref.~\cite{ChengChang2021}, whose text does not mention extinction. The
star-formation rate is linear in halo mass, ${\rm SFR} = A(z)\,M$. A line $X$ emitted by a halo of mass $M$ then has
the luminosity
\begin{equation}
  L(M,z) = 10^{-0.4A_X}\,\frac{{\rm SFR}(M,z)}{K_X} ,
  \label{eq:Lline}
\end{equation}
which is proportional to the halo mass. The conversion coefficients $K_X$ are those adopted in
Refs.~\cite{Gong2017,ChengChang2021}, which follow Ref.~\cite{Kennicutt1998} for H$\alpha$ and \OII{}:
$K_{\Halpha} = 7.9\times10^{-42}$, $K_{\OIII} = 7.6\times10^{-42}$ and $K_{\OII} = 1.4\times10^{-41}$ in units of
$M_\odot\,{\rm yr}^{-1}$ per ${\rm erg\,s^{-1}}$. For H$\beta$, $K_{\Hbeta} = K_{\Halpha}/0.35$~\cite{Gong2017,ChengChang2021}.
The extinctions $A_X$ of Ref.~\cite{Gong2017} are $1.00$, $1.38$, $1.32$ and $0.62$\,mag for H$\alpha$, H$\beta$,
\OIII{} and \OII{}. The line spectra are
\begin{align}
  P^{1h}_{\rm line}(k) &= \int\! dM\,n(M,z)\left[\frac{L(M,z)}{\rho_L}\right]^2 u^2(k|M) ,
  \label{eq:Pline1h}\\
  P^{2h}_{\rm line}(k) &= P_{\rm lin}(k)\Bigl[\int\! dM\,n(M,z)\,\frac{L(M,z)}{\rho_L}\nonumber\\
  &\qquad\qquad\times b(M,z)\,u(k|M)\Bigr]^2 ,
  \label{eq:Pline2h}\\
  P_{{\rm line},\delta}(k) &= \int\! dM\,n(M,z)\,\frac{L(M,z)}{\rho_L}\frac{M}{\bar\rho_m}\,u^2(k|M) \nonumber\\
   &\quad+ P_{\rm lin}(k)\,\mathcal{I}_\delta(k,z)\nonumber\\
   &\qquad\times\int\! dM\,n(M,z)\,\frac{L(M,z)}{\rho_L}\,b(M,z)\,u(k|M) ,
  \label{eq:Plinex}
\end{align}
normalized by the luminosity density $\rho_L$, so that they multiply a kernel without bias. On large scales they
reduce to $b_X^2(z)P_{\rm lin}$ plus Poisson noise, with $b_X(z)$ the luminosity-weighted bias of the line. With one emitter per halo, the one-halo term of
Eq.~\eqref{eq:Pline1h} is the Poisson term of the emitters carrying their profile, as in Eqs.~(2.2) and (2.32) of
Ref.~\cite{MoradinezhadDizgah2022}. The emitting halos are unresolved at every SPHEREx multipole, so this term is white in $\ell$ and equals the Poisson term of Eq.~(24) of Ref.~\cite{ChengChang2021}. The model has no satellite
pairs within a halo and no non-Poissonian corrections to the stochastic term~\cite{MoradinezhadDizgah2022}. The
amplitude $A(z)$ is calibrated before masking, by matching the star-formation-rate density of
Ref.~\cite{HopkinsBeacom2006} over the mass range $10^{8}$--$10^{13}\,h^{-1} M_\odot$ of Ref.~\cite{Gong2017}. The
masking then removes the halos above the mass at which the number density of the masked galaxies is reached, and
their line emission with them. For a catalog complete to $20$\,AB at $1\um$, this threshold mass exceeds
$10^{14}\,M_\odot$ by $z \approx 1.2$, so that the masking affects the lines only below $z \approx 0.8$. Calibrating
$A(z)$ over all halo masses instead is considered as a variant in Sec.~\ref{sec:snrrobust}. The Ly$\alpha$ luminosity is
also proportional to the halo mass, and its amplitude is calibrated to the observed intensity of Ref.~\cite{Pullen2014}.

The model makes three simplifications. First, every line luminosity is proportional to the halo mass, so that the ratios
between the lines are fixed. The gain from the covariance between lines in Sec.~\ref{sec:snr} is therefore an upper
limit, and the summed-line template we use there assumes the line ratios of the model. Second, the two-halo terms are
the bias times the linear matter power spectrum. Third, the redshift-space distortions are linear, without the velocity
dispersion inside halos, and they matter only for the line auto-spectra.

Paper~II~\cite{PaperII} tests the second simplification on simulated maps of SPHEREx and CMB lensing built
from one matter distribution. At $2.2\um$, the measured line auto-spectra exceed a halo model with the measured abundance and bias but a linear
two-halo term by $15$--$30\%$ over $\ell \approx 1000$--$3000$, and the line $\kappa$-cross-spectra exceed it by $1$--$4\%$
at $\ell \approx 1000$ and by $20$--$40\%$ at $\ell \approx 3000$. In those maps, half of the detectability of a line's
cross-spectrum lies below $\ell \approx 400$--$700$, where the cross-spectra agree with the model within their
errors. That paper builds two line models. In the simple model, all halos share the same metallicities; in the metallicity model, the stellar and gas metallicities, and with them the line ratios and the continuum, depend on halo mass and
redshift. We compare the line models in Sec.~\ref{sec:snrrobust}.

\section{Surveys and forecasting method}
\label{sec:surveys}

In this section, we describe the survey configurations we consider and the statistics with which we assess the
detectability of the auto- and cross-spectra.

\subsection{Survey specifications}
\label{sec:specs}

SPHEREx maps the whole sky in 102 spectral channels between $0.75$ and $5\um$~\cite{Bock2025}, with a resolving power
of $36$--$42$ in the four blue bands and of $115$--$135$ in the two red bands. In addition to the all-sky survey, it
observes two deep fields of $100\,{\rm deg}^2$ each, centered near the ecliptic poles at ${\rm Dec} = +66.6^\circ$
and $-71.4^\circ$. We take the noise of each channel map from the sensitivity measured in flight (Fig.~19 of Ref.~\cite{Bock2025}).
In the deep fields, which receive 200 observations per channel, the $1\sigma$ noise $\delta(\nu I_\nu)$ in one $6.2''$
pixel is $0.6$--$4.8\nW$ in all but one channel. In the all-sky survey, which receives four, it is $7.1$ times larger.
The white-noise power of a channel map is then $\sigma_I^2\Omega_{\rm pix}$, with $\sigma_I = \delta(\nu I_\nu)/\nu$
and $\Omega_{\rm pix} = (6.2'')^2$. This noise excludes source confusion, since in an intensity map the unresolved
galaxies are part of the signal.

\begin{figure}[tb]
  \includegraphics{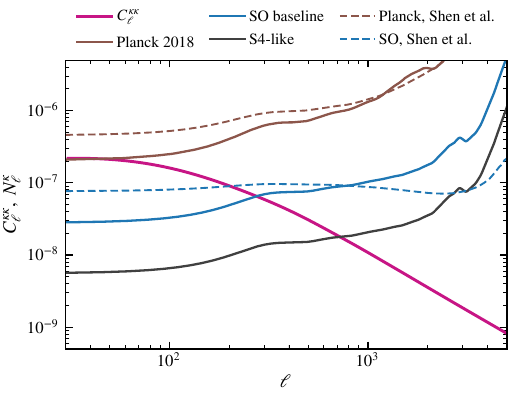}
  \caption{The lensing power spectrum $C_\ell^{\kappa\kappa}$ of the forecast and the reconstruction noise of
  Planck 2018, the SO baseline and the S4-like case (solid). Dashed: the temperature-only noise of
  Ref.~\cite{Shen2025}, used only for the reproductions of their forecasts (Fig.~\ref{fig:shentheory} and
  Appendix~\ref{app:shen}).}
  \label{fig:kappanoise}
\end{figure}

For the CMB-lensing convergence, we model $C_\ell^{\kappa\kappa}$ with the halofit matter power
spectrum~\cite{Smith2003,Takahashi2012} of CLASS~\cite{Blas2011} up to $z = 12$ and the linear one beyond, up to the
last-scattering surface. The cross-spectra of $\kappa$ with the components use the halo-model couplings of
Sec.~\ref{sec:components}, whose two-halo terms carry the linear matter power spectrum. We consider three lensing maps, those of Planck, SO and an S4-like successor, each with its own reconstruction noise, shown
with $C_\ell^{\kappa\kappa}$ in Fig.~\ref{fig:kappanoise} (solid curves). For Planck, this is the minimum-variance
noise of its 2018 lensing reconstruction~\cite{Planck2018lensing}. For SO, it is the baseline minimum-variance noise of
its public noise models,\footnote{\url{https://github.com/simonsobs/so_noise_models}} for an iterative reconstruction from
temperature and polarization~\cite{Ade2019}. The S4-like noise is the SO noise divided by five; it stands for a
hypothetical lower-noise successor rather than a planned experiment. The dashed curves of Fig.~\ref{fig:kappanoise}
are the Planck and SO noise used by Ref.~\cite{Shen2025}, computed with the LensQuEst code of
Ref.~\cite{SchaanFerraro2019} as the $N^{(0)}$ bias of Ref.~\cite{Kesden2003}, in the configuration of Ref.~\cite{Shen2024}:
a reconstruction from temperature alone, for SO with a $1.4'$ beam and $7\,\mu{\rm K}$-arcmin of white noise. Their SO curve
is flatter than the SO baseline, higher at low multipoles and lower at high ones. We use these curves only to reproduce
the forecasts of Ref.~\cite{Shen2025} (Fig.~\ref{fig:shentheory} and Appendix~\ref{app:shen}). The five configurations we consider are listed in
Table~\ref{tab:surveys}. We pair the all-sky survey with the $16\,000\,{\rm deg}^2$ footprint adopted for SO lensing
forecasts~\cite{Ade2019}, or, for Planck, with the $27\,700\,{\rm deg}^2$ its lensing mask keeps ($67\%$ of the
sky)~\cite{Planck2018lensing}.

\begin{table}[tb]
\begin{ruledtabular}
\begin{tabular}{lccc}
configuration & area (deg$^2$) & depth & $N_\ell^{\kappa}$ \\
\colrule
deep $\times$ Planck & $200$ & deep & Planck \\
deep $\times$ SO & $200$ & deep & SO \\
SO footprint $\times$ SO & $16\,000$ & all-sky & SO \\
SO footprint $\times$ S4-like & $16\,000$ & all-sky & SO$/5$ \\
full sky $\times$ Planck & $27\,700$ & all-sky & Planck \\
\end{tabular}
\end{ruledtabular}
\caption{Survey configurations: sky area, SPHEREx depth and lensing reconstruction noise. The deep-field depth
corresponds to 200 observations per channel and the all-sky depth to four. The full-sky area is that kept by the
Planck lensing mask. The lensing noise $N_\ell^{\kappa}$ is that of
Planck~\cite{Planck2018lensing} or of the Simons Observatory (SO)~\cite{Ade2019}; S4-like denotes the SO noise divided
by five.}
\label{tab:surveys}
\end{table}

In the SO-footprint and full-sky configurations, the whole area is taken at the all-sky depth, and the deeper coverage of the deep fields
is not credited. The SPHEREx team builds its own analysis of the background fluctuations on the deep
fields~\cite{Bock2025}. The depth of these fields is set to reach the clustering of the emission from the epoch of reionization, and
their repeated visits allow systematic errors on degree scales to be tested. The joint forecast with Euclid of
Ref.~\cite{EuclidCollab2026} is likewise confined to the deep fields, and Ref.~\cite{Cheng2024} notes that the lower
redundancy of the all-sky survey makes reliable maps harder to construct. A cross-spectrum with lensing weighs these
choices differently. Its continuum part is limited by sample variance, so that the area counts more than the depth,
and noise or systematic errors that do not correlate with $\kappa$ add variance but no bias (Sec.~\ref{sec:snrstat}).
The all-sky survey, however, offers less redundancy, four observations per channel instead of 200, for the null tests that would establish the systematic errors of the maps.

\begin{figure}[tb]
  \includegraphics{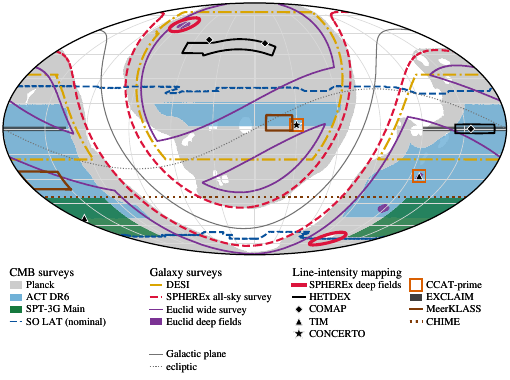}
  \caption{The SPHEREx deep fields, the CMB-lensing maps that could be crossed with them, and the galaxy and line-intensity surveys of the same sky (equatorial Mollweide projection). The Planck~\cite{Planck2018lensing} and ACT DR6~\cite{Madhavacheril2024,Qu2024}
  regions are their released analysis masks, with holes below $20\,{\rm deg}^2$ filled for drawing, and the SO region is the hit map of the nominal scan~\cite{Ade2019}. The dotted CHIME line marks its southern limit, ${\rm Dec} = -40.7^\circ$. The other boundaries follow published
  definitions~\cite{Ge2025,Schlafly2023,EuclidQ12025,Gebhardt2021,Foss2022,Vieira2020,CONCERTO2020,Pullen2022,MeerKLASS2024,CHIME2022}, except three stand-ins: the Euclid wide survey,
  the CCAT-prime fields~\cite{CCATPrime2023}, and the Galactic cut $|b| > 15^\circ$ drawn for the SPHEREx all-sky survey, a stand-in for the
  $75\%$ of the sky its galaxy sample is forecast to keep~\cite{Dore2014}.}
  \label{fig:sky}
\end{figure}

Figure~\ref{fig:sky} places the deep fields on the sky with the lensing maps that could be crossed with them. Only
the Planck map covers them: $96\%$ of the north field and $75\%$ of the
south one, which lies near the Large Magellanic Cloud, masked in the Planck analysis. Neither field lies in the ACT
DR6~\cite{Madhavacheril2024,Qu2024} or SPT-3G~\cite{Ge2025} footprints, or in the $16\,000\,{\rm deg}^2$ footprint
used for SO lensing forecasts, although the nominal SO scan reaches half of the south field. The deep-field
configurations with SO noise assume, as Ref.~\cite{Shen2025}
does, a lensing map that does not yet exist, while the one with Planck noise describes the map that does. The figure
also places the deep fields among the other line-intensity surveys: every published field of HETDEX~\cite{Gebhardt2021}, COMAP~\cite{Foss2022}, TIM~\cite{Vieira2020},
CONCERTO~\cite{CONCERTO2020}, CCAT-prime~\cite{CCATPrime2023}, EXCLAIM~\cite{Pullen2022} and MeerKLASS~\cite{MeerKLASS2024} lies more than $12^\circ$ from the edge of either field, so none overlaps the deep fields.\looseness=-1

\subsection{Signal-to-noise statistics}
\label{sec:snrstat}

We assess the detectability of the cross-spectra with a Fisher analysis. In each multipole bin, the data are the
cross-spectra of all channel maps with the convergence, which we collect in a data vector $\mathbf{d}_\ell$ with one element per
channel, $d_{\ell,i} = C_\ell^{\kappa I_i}$. A bin of width $\Delta\ell$ over a sky area $\Omega$ contains $n_\ell =
\Omega\,\ell\,\Delta\ell/2\pi$ Fourier modes. This count includes a mode and its complex conjugate separately, so that
$n_\ell = 2N_\ell$ with $N_\ell$ the mode count of Paper~II~\cite{PaperII}. For Gaussian fields, the covariance of the measured vector is
then $\Sigma_\ell/n_\ell$, with the covariance per mode~\cite{Knox1995}
\begin{equation}
  \begin{split}
  \Sigma_{\ell,ij} ={}& d_{\ell,i}d_{\ell,j}\\
  &+ \bigl(C_\ell^{\kappa\kappa} + N_\ell^{\kappa}\bigr)
     \bigl(C_\ell^{I_iI_j} + \delta_{ij}\,\sigma_I^2\Omega_{\rm pix}\bigr).
  \end{split}
  \label{eq:crosscov}
\end{equation}
The first term is the product of the signals in channels $i$ and $j$. The second is the product of two total
variances: that of the lensing map, its signal $C_\ell^{\kappa\kappa}$ plus the reconstruction noise $N_\ell^{\kappa}$,
and that of the intensity maps, their signal covariance $C_\ell^{I_iI_j}$ plus the white instrument noise, which is
uncorrelated between channels. Every fluctuation of the intensity maps therefore adds noise to the cross-spectrum, even
when it does not correlate with $\kappa$. The Poisson fluctuations of the line emitters, for example, never enter the
cross-spectrum, but they enter $C_\ell^{I_iI_j}$ and dominate the error bars of the lines.

When the maps are filtered along the LoS, the projection of Eq.~\eqref{eq:projection} acts on both the data
and their covariance, $\tilde{\mathbf{d}}_\ell = \mathsf{H}\mathbf{d}_\ell$ and $\tilde\Sigma_\ell =
\mathsf{H}\Sigma_\ell\mathsf{H}^{\sf T}$, which keeps the correlations that the filter induces between channels.
The filtered covariance is therefore positive semi-definite by construction, and its only null directions are the
$2m+1$ removed modes. We use its pseudo-inverse $\tilde\Sigma_\ell^{+}$, the inverse on the directions that the filter
keeps. Without filtering, it is the ordinary inverse.

The signal covariance $C_\ell^{I_iI_j}$ in Eq.~\eqref{eq:crosscov} can be chosen in two ways, which answer different
questions. If it contains only the component of interest, the result is the detectability of that component in maps
from which every other component has been perfectly removed. This is the convention of Ref.~\cite{Shen2025}, and we
refer to it as the isolated case. If it contains all components, as the measured maps do, the other components add
noise, and their own cross-spectra with $\kappa$ have to be fitted jointly with that of the component of interest. We
refer to this as the joint fit. In the isolated case, the signal-to-noise ratio (SNR) is that of the cross-spectrum against
zero, summed over multipole bins. In the joint fit, we multiply the cross-spectrum $\mathbf{d}^{(a)}_\ell$ of each component $a$
by an amplitude $\mathcal{A}_a$, of fiducial value one, and compute the Fisher matrix of the amplitudes. The two statistics are
\begin{equation}
  \begin{aligned}
  {\rm SNR}^2 &= \sum_\ell n_\ell\, \tilde{\mathbf{d}}_\ell^{\sf T}\,
                 \tilde\Sigma_\ell^{+}\,\tilde{\mathbf{d}}_\ell,\\
  \mathcal F_{ab} &= \sum_\ell n_\ell\, \tilde{\mathbf{d}}^{(a){\sf T}}_{\ell}\,
                 \tilde\Sigma_\ell^{+}\,\tilde{\mathbf{d}}^{(b)}_{\ell},
  \end{aligned}
  \label{eq:crosssnr}
\end{equation}
and the detectability of component $a$ in the joint fit, marginalized over the amplitudes of all other components, is
$\mathcal{A}_a/\sigma(\mathcal{A}_a) = [(\mathcal F^{-1})_{aa}]^{-1/2}$. We sum over $\ell = 36$--$5000$. To assess how much the
forecast depends on the smallest scales, we also stop the sum at $\ell_{\rm max} = 3000$ (Sec.~\ref{sec:snrsurvey}).

For the auto-spectrum, the data are the channel covariance itself. With $S^{(a)}_\ell$ the channel covariance of
component $a$ and $C_\ell = \sum_a S^{(a)}_\ell + {\rm diag}(\sigma_I^2\Omega_{\rm pix})$ the total, the Fisher
information on the amplitude of a Gaussian signal covariance gives
\begin{equation}
  \begin{aligned}
  {\rm SNR}^2 &= \sum_\ell \frac{n_\ell}{2}\,
     {\rm Tr}\bigl[(\tilde S_\ell\,\tilde C_\ell^{+})^2\bigr],\\
  \mathcal F_{ab} &= \sum_\ell \frac{n_\ell}{2}\,
     {\rm Tr}\bigl[\tilde C_\ell^{+}\tilde S^{(a)}_{\ell}\,
                   \tilde C_\ell^{+}\tilde S^{(b)}_{\ell}\bigr],
  \end{aligned}
  \label{eq:autosnr}
\end{equation}
where the tilde again denotes the filtered matrices, $\tilde S_\ell = \mathsf{H}S_\ell\mathsf{H}^{\sf T}$ and
$\tilde C_\ell = \mathsf{H}C_\ell\mathsf{H}^{\sf T}$. As for the cross-spectrum, restricting $C_\ell$ to one component and the noise gives
the isolated detectability of that component, while the second expression is the joint fit of all components, with
the white-noise level fitted as an additional template.

Since a Fisher statistic is unchanged by any invertible transformation of the data, a smooth (apodized) high-pass filter would leave these forecasts unchanged. Only a filter that removes directions from the data, such as the projection of Eq.~\eqref{eq:projection}, changes them.

Finally, the local foregrounds (zodiacal light, diffuse Galactic light and stars) are far brighter than
the extragalactic signal, but statistically independent of the large-scale structure. Their
cross-correlation with $\kappa$ vanishes in the mean, so they bias no cross-spectrum and enter only its
covariance, as Ref.~\cite{ChengChang2021} notes for galaxy tracers. We include the diffuse Galactic light as a variant of the covariance in
Sec.~\ref{sec:snrrobust}. The fluctuations of the zodiacal light are bounded far below the signal~\cite{Zemcov2014},
and the residual of faint stars~\cite{Crill2025} is not modeled here. The extragalactic continuum is a different case.
It is smooth in frequency, like the local foregrounds, but traces the large-scale structure along the lightcone and
correlates with $\kappa$, so any part of it that survives the filter remains in the mean of the measured
cross-spectrum.\looseness=-1

\section{The unfiltered cross-spectra}
\label{sec:unfiltered}

In this section, we present the cross-spectra of the EBL components with the CMB-lensing convergence before any
filtering along the LoS, and the main uncertainties of these predictions. Unless stated otherwise, we show the spectra
at $\lamobs = 2.2\um$.\looseness=-1

\begin{figure*}[p]
\setlength{\abovecaptionskip}{4pt}%
\centering
\includegraphics{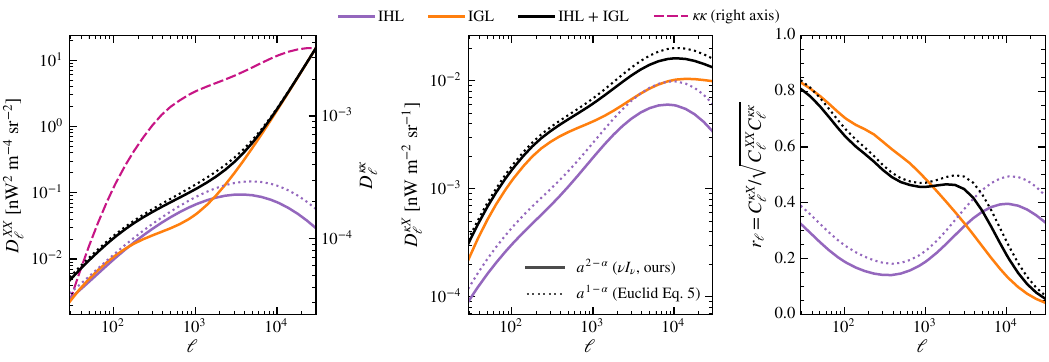}
\caption{The IGL and IHL at $\lamobs = 2.2\um$, unfiltered. Left: the auto-spectra $D_\ell^{XX}$, with the
dimensionless $D_\ell^{\kappa\kappa}$ on the right-hand axis. Center: the cross-spectra $D_\ell^{\kappa X}$ with the
CMB-lensing convergence. Right: the correlation coefficient $r_\ell$. Solid curves use the IHL kernel with
$a^{2-\alpha}$ adopted in this work, dotted ones the $a^{1-\alpha}$ printed in Ref.~\cite{EuclidCollab2026}. The IGL
auto-spectrum includes its Poisson term; the total omits the cross term between the two components.}
\label{fig:eblkappa}

\vspace{3ex}
\includegraphics{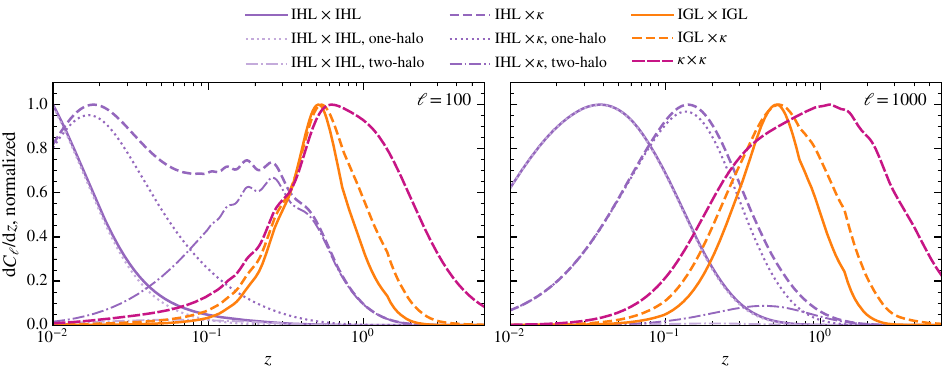}
\caption{Redshift distribution $dC_\ell/dz$ of the auto- and cross-spectra of Fig.~\ref{fig:eblkappa}, at $\ell = 100$
(left) and $\ell = 1000$ (right), and of the CMB-lensing auto-spectrum. The IGL $\times$ IGL curve is the clustering
term, without the Poisson term. Each curve is normalized to its own maximum,
except the one- and two-halo terms of the two IHL spectra, which are divided by the maximum of their total.}
\label{fig:dcdz}

\vspace{3ex}
\includegraphics{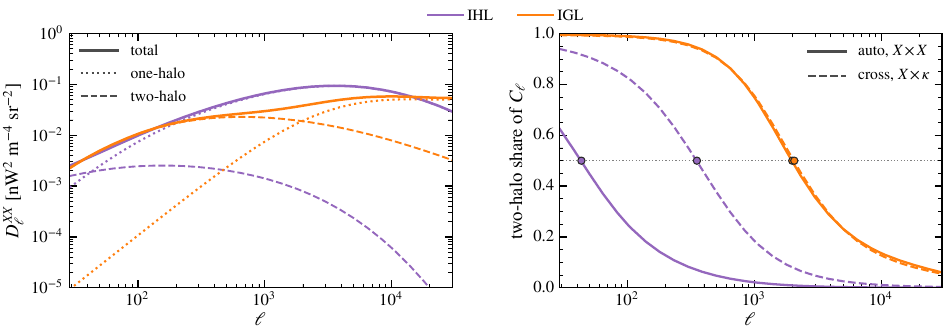}
\caption{The continuum spectra split into one- and two-halo terms. Left: the auto-spectra $D_\ell^{XX}$, total
(solid), one-halo (dotted) and two-halo (dashed). Right: the two-halo share of each $C_\ell$, auto-spectra solid and
cross-spectra with $\kappa$ dashed, with a dot where it crosses one half. The IGL total and its split exclude its Poisson term.}
\label{fig:eblhalo}
\end{figure*}

\subsection{The continuum}
\label{sec:continuum}

To illustrate the relative contributions of the two continuum components, in Fig.~\ref{fig:eblkappa} we show their
auto-spectra (left), their cross-spectra with $\kappa$ (center) and the correlation coefficient $r_\ell =
C_\ell^{\kappa X}/(C_\ell^{XX}C_\ell^{\kappa\kappa})^{1/2}$ (right). Several features are noteworthy in this figure.
First, the IGL dominates the cross-spectrum at every multipole, by a factor of $\XKcrossratlo$--$\XKcrossrathi$ over
$\ell = 30$--$3000$, although the two auto-spectra stay within a factor of $\XKautofactor$ of each other up to $\ell =
2000$. Above $\ell = 2000$, the Poisson term of the unresolved galaxies lifts the IGL auto-spectrum beyond that factor.
Second, the correlation coefficient, which does not depend on the
amplitude of either component, gives the same ordering. Toward high multipoles, the Poisson term of the IGL, which does not correlate with matter, grows from $0.6$ times its clustering power at $\ell =
1000$ to more than three times at $\ell = 3000$, where the two coefficients meet. Third, the IHL cross-spectrum depends on the power of the scale factor in its kernel. With the
$a^{2-\alpha}$ we adopt (Sec.~\ref{sec:components}), it is $\XKIHLapow$ times the one obtained with the $a^{1-\alpha}$
of Ref.~\cite{EuclidCollab2026} at $\ell = 100$.

The ordering of the cross-spectra is set by where in redshift each component emits, relative to the lensing kernel.
To show this, in Fig.~\ref{fig:dcdz} we plot the redshift distribution $dC_\ell/dz$ of each spectrum at $\ell = 100$
(left) and $\ell = 1000$ (right). At $\ell = 100$, the lensing auto-spectrum has its median at $z = \XKzkkmed$ and the
IGL auto-spectrum at $z = \XKzIGLautomed$, so that the unresolved galaxies emit where the lensing
kernel is large. The IHL auto-spectrum, in contrast, has its median at $z = \XKzIHLautomed$ and draws
$\XKzIHLautofa\%$ of its power from below $z = 0.1$, where the lensing kernel is small, while its cross-spectrum lies
in between. As a result, the two auto-spectra are comparable at $\ell = 100$, while the IGL cross-spectrum is several times
the IHL's. At $\ell = 100$, the IHL cross-spectrum also carries both halo terms, which come from different redshifts. Its two-halo term, $\XKtwohalo\%$ of the total, has a median redshift of $z = \XKzIHLcrosstwohalomed$ and its one-halo term of $z \approx \XKzIHLcrossonehalomed$, whereas three quarters of the auto-spectrum is one-halo. At $\ell = 1000$, where each redshift maps onto a scale ten times smaller, the two medians are $\XKzIHLcrosstwohalomedk$ and $\XKzIHLcrossonehalomedk$.
The wiggles of the $\ell = 100$
curves at $z \lesssim 0.4$ are the baryon acoustic oscillations, mapped into redshift by $k = (\ell+\tfrac12)/\chi$.
The multipole dependence of $r_\ell$ follows from the same redshift distributions. In the Limber limit, $r_\ell$ is the overlap of
the two maps, the area under the geometric mean of the redshift distributions of their variance, times the
three-dimensional correlation coefficient of the component with matter averaged over that overlap. The minimum of the
IHL coefficient at $\ell \approx \RSihlellmin$ is a minimum of the overlap, whereas the IGL coefficient falls at high multipoles because its Poisson term decorrelates it from matter.\looseness=-1

The same weighting in redshift explains how the one- and two-halo terms share the multipoles in each of the four spectra,
which we show in Fig.~\ref{fig:eblhalo}. The IGL auto- and cross-spectra are two-halo signals below $\ell \approx
2000$. The IHL cross-spectrum turns one-halo at $\ell \approx \XKtwohaloell$, while the IHL auto-spectrum does so already at $\ell \approx \XKtwohaloautoell$. By probing both halo
terms, the cross-spectrum tests the IHL model more fully than the auto-spectrum of a deep field, which measures mainly
its one-halo term (Sec.~\ref{sec:continuumprobe}). This difference arises from the weight given to each distance. An auto-spectrum weights each distance by $1/\chi^2$, whereas in a cross-spectrum
the lensing kernel, which vanishes linearly at the observer, turns this weight into $1/\chi$. The IHL auto-spectrum
therefore draws on $z < 0.1$, where a given multipole probes scales inside halos, while its cross-spectrum reaches out
to $z \approx 0.3$, where the same multipole probes the clustering of halos.

Since the ordering in $r_\ell$ is set by the redshift overlap, it depends on the lensing field but not on the
amplitudes of the components. CMB lensing weights the matter mostly at $z \approx
0.5$--$4$, where the IGL is bright and the IHL has faded, whereas a cosmic-shear map with nearby sources weights lower
redshifts and selects the IHL instead. In Appendix~\ref{app:shear} we show that, for an approximate Euclid-like
tomographic distribution of sources, the two bins of photometric redshift $z_{\rm ph} \le 0.6$ correlate with the IHL at $r_\ell =
0.62$--$0.85$ and with the IGL at only $0.20$--$0.45$ over $\ell = 100$--$1000$, while sources at $z_{\rm ph} =
1.3$--$2.5$ select the IGL, as CMB lensing does. This reconciles our ordering with the finding of
Ref.~\cite{EuclidCollab2026} that the IHL correlates more strongly than the IGL with Euclid cosmic shear.

\subsection{The emission lines}
\label{sec:lines}

The kernel of an emission line is a single shell. At a given $\lamobs$, a line of rest wavelength $\lambda_{\rm line}$ is seen only from $z = \lamobs/\lambda_{\rm line} - 1$, over the depth of one channel, which we model as a top hat in wavelength. This
has two consequences for the line cross-spectra. First, they do not depend on the resolving power: a thinner shell
raises the kernel and shortens the radial integral by the same factor, so that tripling the resolving power $R$ changes the cross-spectrum
by only $10^{-4}$. The auto-spectra, in contrast, depend on $R$. Second, the Poisson fluctuations of the emitters do not
correlate with $\kappa$ and do not enter the cross-spectrum. They lower the correlation coefficient, to
$\LKtotrlo$--$\LKtotrhi$ for the sum of the four lines at $\LKlam\um$ over $\ell = 30$--$10^4$, and raise the error bars. As discussed in
Sec.~\ref{sec:projection}, the Limber approximation remains accurate for the line cross-spectra, because the smooth
lensing kernel restricts them to modes with $k_\parallel \approx 0$, but not for the line auto-spectra.\looseness=-1

\begin{figure*}[!tb]
\centering
\includegraphics{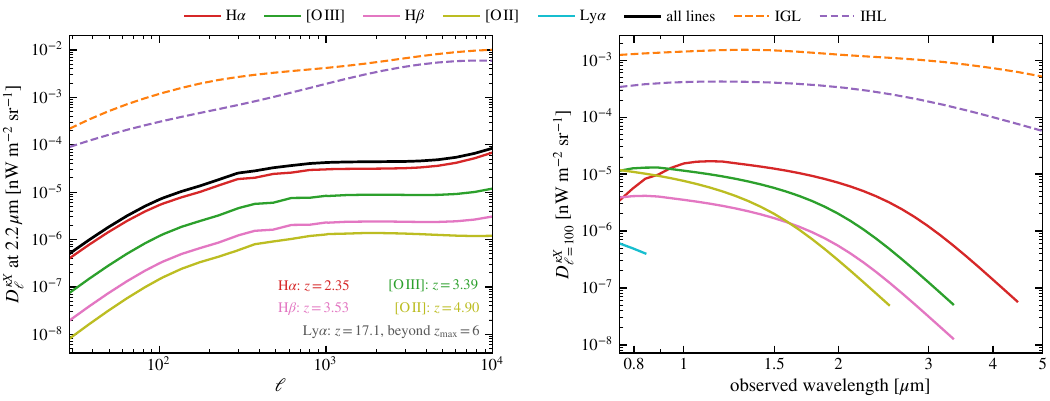}
\caption{Emission-line intensity $\times$ CMB lensing, unfiltered. Left: $D_\ell^{\kappa X}$ at $\lamobs =
\LKlam\um$, one curve per line at the redshift that this wavelength selects, and their sum in black. Right:
$D_{\ell=100}^{\kappa X}$ across the SPHEREx band; each curve ends where its line reaches $z = \LKzmax$, the end of the tabulation of $P(k,z)$ used for this figure, which for Ly$\alpha$ is at $\LKlyalamhi\um$. In both panels, the IGL and IHL cross-spectra are dashed for comparison. The masking,
$20$\,AB at $1\um$, is applied to the lines as to the continuum.}
\label{fig:linekappa}
\end{figure*}

In Fig.~\ref{fig:linekappa} we compare the lines with the continuum: their cross-spectra at $\lamobs = \LKlam\um$
(left), and their amplitude at $\ell = 100$ across the SPHEREx band (right), each beside the IGL and IHL. The lines are
about two orders of magnitude below the continuum. At $\LKlam\um$ and $\ell = 100$ they make up $\LKlinepc\%$ of the
cross-spectrum of the whole EBL, most of it from H$\alpha$, and across the band even H$\alpha$ at its peak stays two
orders of magnitude below the IGL. The two signals accumulate differently along the LoS. At a given wavelength,
the continuum collects the emission of every redshift, whereas a line collects only that of the one shell its channel
selects; in our model, the mean intensity of H$\alpha$ in the channel at $\LKlam\um$ is $0.6\%$ of that of the IGL. For a
SPHEREx map, the continuum is therefore not a small contaminant of a line $\times\,\kappa$ measurement but its
dominant signal. Section~\ref{sec:filtered} asks whether filtering along the LoS changes this.

\subsection{Uncertainties of the predictions}
\label{sec:xkcaveats}

First, the amplitude of the IHL is the least constrained input. It is set by the fraction of the halo light in
diffuse form, which observations constrain poorly.\footnote{We take the value from the joint fit of Ref.~\cite{EuclidCollab2026}; however, we note a discrepancy
between our auto-spectra and their Fig.~2 (Appendix~\ref{app:validation}). With the IHL power of that figure, five
times lower than ours, the summed lines would reach $9.2\sigma$ rather than $7.4\sigma$ over the SO footprint, and
the IHL $40\sigma$ rather than $63\sigma$. Computed in their conventions, our IGL is $\EFtwoIGLavg$ times theirs at every $\ell$, a factor close to $1/h^3$ for their
$h = 0.67$.}
A change in this amplitude rescales the IHL auto-spectrum quadratically and the cross-spectrum linearly, and leaves $r_\ell$ unchanged,
so that the ordering in $r_\ell$ of Sec.~\ref{sec:continuum} is unaffected. The IGL would cease to dominate the cross-spectrum at $\ell = 100$
only if the IHL emissivity were about four times that of our model.

Second, the IHL auto-spectrum of a deep field is set by a handful of halos. It draws $\XKzIHLautofa\%$ of its power from
$z < 0.1$, where a $100\,\mathrm{deg}^2$ field contains few massive halos. For the mass function of Ref.~\cite{ShethTormen1999}, it holds about $\XKnclusterlo$ halos above $10^{14}\,M_\odot$ at $z < 0.05$ and $\XKnclustermid$ at $z = 0.05$--$0.1$, while the one-halo term comes from halos around $\XKihlmedmass\,M_\odot$. The auto-spectrum
measured in one field can therefore differ from the ensemble average by a factor of order unity, an estimate from the
halo counts rather than from simulations. The cross-spectrum is much less exposed, since only $\XKzIHLcrossfa\%$ of it comes
from $z < 0.1$, and raising the lower redshift limit from $0.01$ to $0.1$ lowers it by
$\XKzcrossdrop\%$, against a factor of $\XKzautofac$ for the auto-spectrum.

Third, the cross-spectrum of the whole EBL with $\kappa$ is exactly the sum of those of its components, since the
projection is linear in the emission, but its auto-spectrum is not. The IGL and the IHL occupy the same halos, and the
line emitters are the galaxies whose starlight makes up the IGL, so the cross terms between components, which our baseline omits, are positive. The sum of the auto-spectra of the components is therefore a lower bound on the total,
and the corresponding $r_\ell$ an upper bound. In Sec.~\ref{sec:snrrobust} we include these cross terms as a variant of
the covariance.\looseness=-1

\section{The filtered spectra}
\label{sec:filtered}

In this section, we examine how a filter along the LoS affects the auto-spectra and the cross-spectra with $\kappa$.
We ask whether it removes the smooth continuum while sparing the thin shell a line occupies in
one channel, and so reveals the lines, whose cross-spectrum is two orders of magnitude below that of the continuum
(Sec.~\ref{sec:lines}).

We apply the projection of Eq.~\eqref{eq:projection} to the channel profiles and covariances of
Eq.~\eqref{eq:filtobs}, for the IGL, the IHL, H$\alpha$, \OIII{} and \OII{}. We summarize a filtered auto-spectrum by its
band-rms amplitude, $\langle C_{ii}\rangle^{1/2}$ over the channels, and a filtered cross-spectrum by the rms of its
profile. The surviving fraction is the ratio of the filtered to the unfiltered value. For a smooth component, the
surviving part comes mostly from the edges of the band, so its surviving fraction depends on details of the filter: whether
it acts on $I_\nu$ or $\nu I_\nu$, which channels it acts on, and its exact form (Appendix~\ref{app:filters}). Our
values are for the projection applied to $I_\nu$ on the 102 SPHEREx channels.

\begin{figure}[t]
\includegraphics{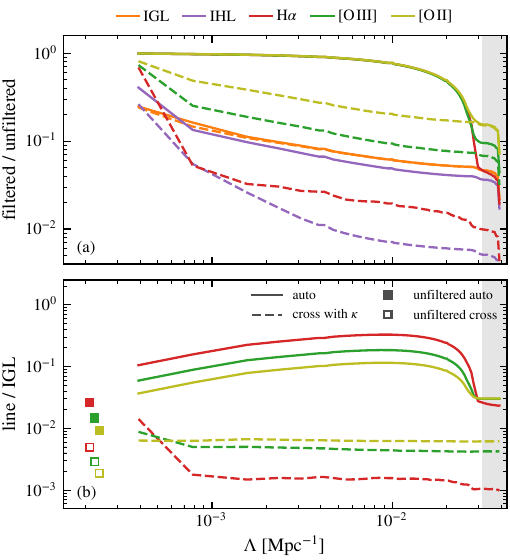}
\caption{What the filter along the LoS leaves of each component at $\ell = 200$, for the projection of
Eq.~\eqref{eq:projection} applied to the specific intensity $I_\nu$ on the 102 SPHEREx channels. (a) Surviving fraction of the band-rms amplitude, filtered over unfiltered, against the filter scale
$\Lambda$, for the auto-spectra (solid) and the cross-spectra with $\kappa$ (dashed). (b) Ratio of each line to the IGL
for the same statistics, with the unfiltered ratios as squares at the left. Each curve starts at $m = 0$, where only
the channel mean is removed. The shaded region lies beyond the Nyquist wavenumber of one channel at $2.2\um$, $k_\parallel
\approx 0.03\Mpcinv$, where the projection runs out of channel directions (Sec.~\ref{sec:filter}).}
\label{fig:filtsurv}
\end{figure}

To illustrate what the filter leaves of each component, in Fig.~\ref{fig:filtsurv} we show the surviving fraction
against the filter scale $\Lambda$ at $\ell = 200$, together with the ratio of each line to the IGL. In
Table~\ref{tab:survival} we list the surviving fractions over $\ell = 200$--$1000$ at $m = 0$ and at $\Lambda =
0.01\Mpcinv$, a radial wavelength of about $600$\,Mpc.

The auto-spectra behave as expected from the intensity-mapping literature [Fig.~\ref{fig:filtsurv}(a), solid
  curves]. The line auto-spectra keep most of their amplitude, since a channel shell about $100$\,Mpc deep carries its power up to $k_\parallel \approx 0.03\Mpcinv$, well above $\Lambda$. The continuum
  auto-spectra keep only a few percent, and most of the loss occurs already when the channel mean is removed
  (Table~\ref{tab:survival}). Every continuum channel integrates the same lightcone, so that the continuum varies
  across channels as smoothly as its spectrum, apart from the edges of the band. The line-to-IGL ratio of the
  auto-spectra therefore rises by an order of magnitude, but never reaches unity [Fig.~\ref{fig:filtsurv}(b)]. In the
auto-spectrum, filtering helps separate the lines from the continuum, but does not isolate them.

In the cross-spectra, the lines do not separate from the continuum [Fig.~\ref{fig:filtsurv}(a), dashed curves]. Removing the channel mean lets
  the lines gain on the IGL, yet the line-to-IGL ratio of the cross-spectra stays below $0.015$ at every $\Lambda$
  [Fig.~\ref{fig:filtsurv}(b)]. At $\Lambda = 0.01\Mpcinv$, H$\alpha$ falls further behind the IGL than when
  unfiltered, while \OIII{} and \OII{} gain (Table~\ref{tab:survival}).
\begin{table}[!t]
\fitwidth{\begin{tabular}{lccccc}
\toprule
 & \multicolumn{2}{c}{auto [\%]} & \multicolumn{2}{c}{cross with $\kappa$ [\%]} & line/IGL cross \\
 & $m = 0$ & $\Lambda = 0.01$ & $m = 0$ & $\Lambda = 0.01$ & unfiltered $\to$ $0.01$ \\
\midrule
IGL & $25$--$26$ & $6.2$--$6.4$ & $24$ & $5.9$--$6.2$ &  \\
IHL & $39$--$41$ & $4.3$--$4.9$ & $26$--$29$ & $0.33$--$0.71$ &  \\
H$\alpha$ & $99$ & $78$--$80$ & $67$--$69$ & $1.7$--$2.0$ & $0.0050 \to 0.0016$ \\
\OIII & $99$ & $77$--$79$ & $71$--$73$ & $6.0$--$9.4$ & $0.0029 \to 0.0044$ \\
\OII & $99$ & $78$--$80$ & $78$--$81$ & $16$--$20$ & $0.0019 \to 0.0063$ \\
\bottomrule
\end{tabular}
}
\caption{Surviving fraction of the band-rms amplitude of each component over $\ell = 200$--$1000$ (the power fraction
is its square), for the projection applied to $I_\nu$ with only the channel mean removed ($m = 0$) and at $\Lambda =
0.01\Mpcinv$. The last column gives the line-to-IGL ratio of the cross-spectra at $\ell = 200$, unfiltered and at
$\Lambda = 0.01\Mpcinv$.}
\label{tab:survival}
\end{table}
  A line occupies a thin shell along the LoS,
  but this sharpness matters only when the shell is correlated with something equally sharp. In the auto-spectrum, the
  shell is correlated with itself. In the cross-spectrum, it is correlated with $\kappa$, whose kernel is smooth over
  gigaparsecs. The cross-spectrum of a line therefore varies from channel to channel as smoothly as that of the
  continuum, following the slow evolution of the emission and of the lensing kernel with redshift, and the filter
removes both alike.

The power that survives the filter comes from the edges of the band. At $\Lambda = 0.01\Mpcinv$, $84$--$96\%$ of the surviving cross
  power of the IGL, the IHL, \OIII{} and \OII{} lies in the outer fifth of the channels. The edges of the band at $0.75$
  and $5\um$ are steps, broadband in $k_\parallel$, and a profile that is still high where the band cuts it survives
  through them. This survival is set by the edges of the band rather than by the evolution of the emission. The \OIII{} and \OII{}
  cross-spectra peak near the blue edge, which cuts them at $z \approx 0.5$ and $1$ while they are still high, and
  survive through it. Tapering the ends of the band before filtering removes most of what they keep
  (Appendix~\ref{app:filters}). The H$\alpha$ cross-spectrum, in contrast, falls off inside the band at both ends: it
  keeps the least of the lines, but most of what it keeps comes from the evolution of its profile within the band. Since smooth
  modes remove a step only slowly, the dashed curves of Fig.~\ref{fig:filtsurv}(a) flatten at large $\Lambda$. The line auto-spectra fall from about
  $0.5$ to $0.05$--$0.16$ between $\Lambda \approx 0.02$ and $0.03\Mpcinv$, before the shaded region, which lies beyond the
  Nyquist wavenumber of one channel at $2.2\um$ (Sec.~\ref{sec:filter}). Inside it, the curves flatten, and every curve
  drops at the last point, $\Lambda \approx 0.04\Mpcinv$, where the filter removes the last channel directions.\looseness=-1

Separating the lines from the continuum in the cross-spectrum with lensing therefore requires another approach. In
Sec.~\ref{sec:snr}, we show that fitting the spectra of all components together provides it.

This result does not contradict Ref.~\cite{Shen2025}, whose forecast concerns single-line experiments and sets the
spectrally smooth extragalactic background aside (their Table~I and footnote~1). With a single line and no continuum, their
forecast does not need to separate the two.

Reference~\cite{Shen2025} recommended against the Limber approximation for filtered maps. For the filtered
cross-spectra with $\kappa$, we find it accurate to $0.7\%$ or better, and to $0.23\%$ for the lines. The filter acts
linearly on the cross-spectrum of each channel, after the integral along the LoS, and that cross-spectrum is accurate
in the Limber approximation (Sec.~\ref{sec:projection}). It fails only for the line auto-spectra, as it
does without filtering (Appendix~\ref{app:limber}). Appendix~\ref{app:shen} traces the deviation of Ref.~\cite{Shen2025} to the
discretization of their two integrals and, for COMAP, to the cut of their kernel at the edges of the redshift window.\looseness=-1

\section{Detectability}
\label{sec:snr}

In this section, we assess the detectability of the EBL components in the survey configurations of
Table~\ref{tab:surveys}, and how the line detection depends on the shape of the continuum across the channels and on
the modeling. We compute these forecasts from the halo model of Sec.~\ref{sec:components}, and in Paper~II~\cite{PaperII}
we apply the same statistics to simulated SPHEREx and CMB-lensing maps.

\subsection{Each component alone}
\label{sec:snrall}

\begin{figure*}[tb]
  \centering
  \includegraphics{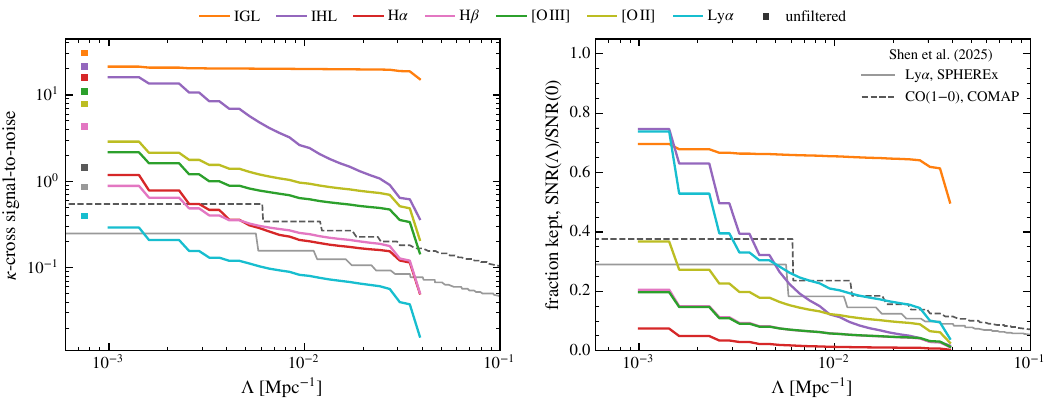}
  \caption{Signal-to-noise ratio of the $\kappa$-cross-spectrum of each SPHEREx component alone in the map
  [Eq.~\eqref{eq:crosssnr}, isolated case], in the two deep fields with SO lensing noise, summed over $\ell =
  36$--$5000$. Left: against the filter scale $\Lambda$ of the projection, with the unfiltered values as squares at
  the left edge. Right: the fraction of the unfiltered value that survives. Gray: the forecasts of
  Ref.~\cite{Shen2025} for SPHEREx Ly$\alpha$ over $200\,{\rm deg}^2$ (solid) and COMAP CO(1--0) over $12\,{\rm
  deg}^2$ (dashed), rebuilt with their model, noise and filter in the Limber and noise-dominated approximations,
  without their corrections for these approximations (their Figs.~13--14; Appendix~\ref{app:shen}), with the unfiltered
  values of the rebuilt forecasts as gray squares. Their published unfiltered values are $0.87$ (SPHEREx Ly$\alpha$) and $1.24$
  (COMAP). Our
  curves end where the projection has removed every channel direction, while theirs continue because they filter
  before binning (Sec.~\ref{sec:filter}).}
  \label{fig:shentheory}
\vspace{1.5ex}%
\end{figure*}

To compare the components on the same footing as Ref.~\cite{Shen2025}, in Fig.~\ref{fig:shentheory} we show the
signal-to-noise ratio of each component alone in the map, against the filter scale. Unfiltered (colored squares), the
IGL and the IHL reach signal-to-noise ratios of about $30$ and $20$ in the two deep fields, and the lines follow in the
order of their intensity, from H$\alpha$ at $16$ to Ly$\alpha$ at $0.4$. Reference~\cite{Shen2025} anticipated that H$\alpha$ is the
better target (their footnote~8); we find it about forty times as detectable as Ly$\alpha$, and its detectability comes
from $z \approx 1$--$2.5$, near the peak of the lensing kernel. In every model we consider, Ly$\alpha$ is not detectable. Our Ly$\alpha$ emission,
calibrated to Ref.~\cite{Pullen2014}, is about half as bright in intensity times bias as that of
Ref.~\cite{Shen2025}.

The filter costs the cross-spectrum a large and uneven fraction of its signal-to-noise ratio. At $\Lambda =
0.01\Mpcinv$, the IGL keeps about two thirds, the IHL about a tenth, and the lines between $1$ and $21\%$.
Ly$\alpha$ and \OII{} keep the most, because their profiles are still high where their range of channels ends, and
H$\alpha$, whose profile falls off inside the band, keeps the least (Sec.~\ref{sec:filtered}). For the IGL, the filter
removes the sample variance along with the signal, and so costs far less in signal-to-noise ratio than in signal, of which
about $6\%$ survives. The variance of the lines is nearly diagonal across channels and survives the filter, so that
filtering widens the lead of the continuum.

The forecasts of Ref.~\cite{Shen2025} for SPHEREx Ly$\alpha$ and COMAP CO(1--0) (gray curves) keep a fifth to a
quarter of their signal-to-noise ratio at
$\Lambda = 0.01\Mpcinv$, more than our lines other than Ly$\alpha$. This is due to the redshift ranges of the lines rather than to the filter. Each of their lines spans a narrow range, about $1000$\,Mpc deep, which cuts its profile
off at both ends. Our Ly$\alpha$ occupies the same range, $z = 5.2$--$8$, and keeps as much as theirs, although
we filter all 102 channels. In Appendix~\ref{app:shen} we reproduce their forecasts from their model and
discuss their assessment of the Limber approximation.\looseness=-1

\subsection{All components fitted jointly}
\label{sec:snrsurvey}

\begin{table*}[tb]
  \setlength{\tabcolsep}{12.5pt}%
  \fitwidth{\begin{tabular}{lcccc}
\toprule
 & \multicolumn{4}{c}{$\kappa$-cross, all components fitted at once} \\
survey & IGL & IHL & H$\alpha$ & summed lines \\
\midrule
deep fields $\times$ Planck & 9.2 / 6.6 / 0.93 & 3.3 / 3.0 / 0.23 & 0.36 / 0.35 / 0.02 & 0.88 / 0.75 / 0.03 \\
deep fields $\times$ SO & 23 / 17 / 2.2 & 9.8 / 9.0 / 0.65 & 0.81 / 0.79 / 0.04 & 2.0 / 1.6 / 0.08 \\
SO footprint $\times$ SO & 173 / 84 / 8.2 & 63 / 61 / 1.7 & 3.0 / 2.8 / 0.10 & 7.4 / 7.3 / 0.28 \\
SO footprint $\times$ S4-like & 276 / 128 / 12 & 110 / 106 / 2.8 & 4.0 / 3.7 / 0.13 & 11 / 11 / 0.38 \\
full sky $\times$ Planck & 96 / 50 / 5.3 & 31 / 29 / 0.92 & 2.0 / 1.9 / 0.08 & 4.6 / 4.5 / 0.17 \\
\midrule
 & \multicolumn{4}{c}{auto-spectrum, all components and the noise level fitted} \\
 & IGL & IHL & H$\alpha$ & \\
\midrule
deep fields & 373 & 269 & 47 & \\
SO footprint (all-sky depth) & 2124 & 1817 & 20 & \\
\bottomrule
\end{tabular}
}
\caption{Detectability with every component in the map. Top: the $\kappa$-cross-spectrum with all components fitted
  jointly and marginalized, for the IGL, the IHL and H$\alpha$ (all seven templates fitted) and for the summed lines
  (one template for the five lines, beside the two continuum templates). Each cell gives the value unfiltered / with
  only the channel mean removed ($m = 0$) / after the projection at $\Lambda = 0.01\Mpcinv$. Bottom: the
  auto-spectrum, unfiltered, with every component and the white-noise level fitted. The survey configurations are
  those of Table~\ref{tab:surveys}. Of the individual lines, only H$\alpha$ is listed: with all components fitted, each
  of the others stays below $1.6\sigma$ in the $\kappa$-cross-spectrum in every configuration. Appendix~\ref{app:surveytable} gives every component,
  alone and fitted jointly.}
\label{tab:survey}
\end{table*}

A measured channel map contains the continuum, whose cross-spectrum with $\kappa$ is a hundred times that of a line
and whose sample variance enters the error bar of every channel. In Table~\ref{tab:survey} we therefore fit the
amplitudes of all components at once. This degrades the detectability of the continuum only mildly, but that of the lines drastically. In the deep
fields with SO lensing noise, the IGL and the IHL lose a quarter and a half of their signal-to-noise ratios, respectively, while that
of H$\alpha$ falls twentyfold. Most of this loss, a factor of eight, comes from the sample variance of the continuum in
the line channels, and the rest from the degeneracy between the line templates, strongest between H$\beta$ and
\OIII{}, whose amplitudes correlate at $-0.96$ (Sec.~\ref{sec:snrauto}). The IGL and the IHL, in contrast, separate
well: the correlation of their fitted amplitudes does not exceed $0.38$.

Fitting one template for the five lines, with the line ratios of the model, lifts the degeneracy between them: the
summed lines reach $7.4\sigma$ over the SO footprint, against $3.0\sigma$ for H$\alpha$ in the seven-template fit.
These values take the shape of the continuum templates across the channels as exact, an assumption we examine in
Sec.~\ref{sec:snrshape}.

\begin{table*}[tb]
  \setlength{\tabcolsep}{17pt}%
  \fitwidth{\begin{tabular}{lccc}
\toprule
continuum's shape & deep fields & SO footprint & S4-like \\
\midrule
continuum shape known & 2.0 & 7.4 & 10.7 \\
one / two / three polynomial terms per template & 1.3 / 1.1 / 0.9 & 5.9 / 4.8 / 3.4 & 8.5 / 6.8 / 4.4 \\
8 / 20 parameters, from the $\kappa$-cross alone & 0.7 / 0.4 & 2.3 / 1.4 & 3.1 / 2.1 \\
8 / 20 parameters, priors at the step & 1.1 / 0.7 & 3.1 / 2.0 & 4.1 / 2.9 \\
8 / 20 parameters, calibrated by the continuum auto-spectra & 2.0 / 1.9 & 7.4 / 7.3 & 10.6 / 10.6 \\
three noise-weighted principal components removed & 1.8 & 5.7 & -- \\
\bottomrule
\end{tabular}
}
\caption{Signal-to-noise ratio of the summed-line $\kappa$-cross-spectrum, unfiltered, for each treatment of the
  shape of the continuum across the channels. Known: the IGL and IHL templates are exact, as in
  Table~\ref{tab:survey}. Polynomial terms: each continuum template is multiplied by a polynomial in $\ln\lambda$ of
  degree one, two or three with free coefficients. Parameters: the templates depend on eight or twenty physical
  parameters of the IGL and IHL (Sec.~\ref{sec:snrshape}), fitted from the $\kappa$-cross-spectrum alone, without
  priors or with Gaussian priors as wide as their finite-difference steps, or calibrated by the auto-spectra of the
  continuum. Principal components: the three leading principal components of the observed channel covariance, with
  each channel weighted by its inverse noise, are removed, and the summed lines are fitted beside the two continuum
  templates (Sec.~\ref{sec:channelmodes}). The columns are the deep
  fields and the SO footprint with SO lensing noise, and the SO footprint with the S4-like noise
  (Table~\ref{tab:surveys}); a dash marks a case not computed.}
\label{tab:ladder}
\end{table*}

Filtering does not help. It removes the variance of the continuum together with the cross-spectrum of the lines
(Sec.~\ref{sec:filtered}), and at $\Lambda = 0.01\Mpcinv$ every value of Table~\ref{tab:survey} falls by $90\%$ or
more, even that of the IGL. Removing only the channel mean costs H$\alpha$ alone a third of its
signal-to-noise ratio, but barely changes its joint fit, which had already lost that part to the continuum. The lines can therefore be detected only by fitting the continuum.\looseness=-1

Because the detection is limited by sample variance, the area of the survey matters more than its depth or the
lensing noise. In the deep fields, lowering the lensing noise fivefold or deepening the SPHEREx maps by one magnitude
raises H$\alpha$ in the joint fit only to about one (Appendix~\ref{app:surveytable}), whereas the all-sky survey raises it to
three. The continuum gains even more, about sevenfold, because the maps remain dominated by its signal at the all-sky
depth. The Planck map, the only lensing map that currently covers the deep fields (Sec.~\ref{sec:specs}), gives
signal-to-noise ratios there two to three times lower than SO noise would. Over the sky that its mask keeps, it gives less than the
SO footprint for every component, since its reconstruction noise is about thirteen times that of SO at $\ell \approx
1000$. Deriving the SPHEREx noise from its point-source depths instead, with the method of Ref.~\cite{Shen2025}, who use one
depth, $m_{\rm AB} = 22$, for every channel, or crediting the deep fields with their fainter polar sky changes the
detectability of the lines by at most $11\%$.

We note two caveats to these values. First, the covariance is Gaussian, as in Ref.~\cite{Shen2025}, without the
connected trispectrum or the super-sample covariance. These matter most for the continuum, whose signal-to-noise
ratio is limited by sample variance and accumulates at high multipoles. Half of the ${\rm SNR}^2$ of the IGL and IHL cross-spectra lies above $\ell \approx 750$ and $820$, respectively, where the one-halo term dominates the IHL
cross-spectrum and is a growing part of the IGL's, reaching half at $\ell \approx 2000$. Stopping the sum at $\ell = 3000$ instead of $5000$ lowers the detections of the summed lines and of
the IGL by at most $2\%$ and those of the IHL by $6\%$. The size of the non-Gaussian terms, however, has to be
assessed before the continuum values are quoted as detections. Second, the local foregrounds enter only the
covariance (Sec.~\ref{sec:snrstat}), but the CIB that leaks into temperature-based lensing
estimators~\cite{Osborne2014,vanEngelen2014} correlates with the SPHEREx maps, and so biases the mean of the
cross-spectrum. Bias-hardened or polarization-only estimators mitigate this bias, but we have not evaluated its size here.

\subsection{The spectral shape of the continuum}
\label{sec:snrshape}

The joint fit of Sec.~\ref{sec:snrsurvey} takes the shapes of the IGL and IHL templates across the channels as
exact. Since the cross-spectrum of a line is less than a percent of that of the IGL in a channel (Sec.~\ref{sec:lines}), a wrong
template biases the lines even when its error is small and smooth. Interpolating the normalization of the luminosity
function~\cite{Helgason2012} linearly rather than in its logarithm changes the IGL template by $2.1\%$ across the
band, almost all of it smoothly. The free amplitudes of the IGL and IHL absorb only part of this change, and the rest,
comparable to the cross-spectrum of the lines, shifts the summed-line amplitude by $41\%$, or $3.0\sigma$ over the SO
footprint. We therefore consider three scenarios, which we expect to bracket what an analysis could achieve: the
shape is known, it is fitted from the $\kappa$-cross-spectrum alone, or it is calibrated by the auto-spectra of the
continuum. In Table~\ref{tab:ladder} we list the summed-line detection in each scenario, together with intermediate
treatments.

Fitted from the cross-spectrum alone, the shape depends on eight physical parameters: the six of the luminosity function (common to all its bands), the masking depth and the redshift evolution of the IHL. Freeing the normalization in each band of Ref.~\cite{Helgason2012} as well gives twenty. These parameters absorb the template error above, but the detection falls to
$2.3\sigma$ with eight of them and $1.4\sigma$ with twenty. Together, they can build a pattern across the channels
that resembles the summed lines, and a $14\%$ measurement of the line amplitude requires that pattern to be known to
about $0.09\%$. Reference~\cite{Helgason2012} publishes its evolution parameters without uncertainties, and even
priors of $0.1$\,mag on the characteristic magnitude and $5\%$ on every normalization recover only $2$--$3\sigma$.

\begin{table*}[tb]
  \fitwidth{\begin{tabular}{lcccccc}
\toprule
 & UniverseMachine & baseline & uniform dust & no dust & simple model & metallicity model \\
\midrule
deep fields $\times$ SO, summed lines & 1.1 & 2.0 & 1.8 & 4.6 & 8.1 & 5.3 \\
deep fields $\times$ SO, H$\alpha$ & 0.49 & 0.80 & 0.63 & 1.5 & 5.7 & 3.9 \\
SO footprint $\times$ SO, summed lines & 4.2 & 7.3 & 6.4 & 21 & 45 & 23 \\
SO footprint $\times$ SO, H$\alpha$ & 1.7 & 3.0 & 2.3 & 6.8 & 30 & 17 \\
\bottomrule
\end{tabular}
}
\caption{Signal-to-noise ratio of the $\kappa$-cross-spectrum per line model, unfiltered: the summed lines with the
  continuum fitted, and H$\alpha$ with all components fitted, in the deep fields and over the SO footprint, with SO
  lensing noise. Models: the star-formation rate per halo of UniverseMachine~\cite{Behroozi2019} with the
  coefficients of Ref.~\cite{Kennicutt1998}, consistent with the Ly$\alpha$ model of Ref.~\cite{Shen2025}; the
  baseline, with the dust of Ref.~\cite{Gong2017}; a uniform $1.30$\,mag of dust, the effective extinction inferred
  from the line intensities of Ref.~\cite{ChengChang2021}; no dust; and the lines of the simple and metallicity models
  of the simulated maps of Paper~II~\cite{PaperII}, computed with the Flexible Stellar Population Synthesis
  code~\cite{Conroy2009}. The lines are evaluated in the Limber approximation here.}
\label{tab:variants}
\end{table*}

The channel maps, however, measure the shape of the continuum directly. Every parameter that moves its cross-spectrum
moves its channel covariance too, and the auto-spectra are detected at signal-to-noise ratios in the hundreds
(Sec.~\ref{sec:snrauto}). In the calibrated scenario, we add the channel auto-spectra to the cross-spectrum as
independent data, with the same parameters acting on both. The amplitudes of all auto-spectra, including those of the lines, and the noise level are left free, so that the auto-spectra inform only the shape of the continuum. The
detection is then restored to $7.4\sigma$ over the SO footprint with eight parameters and $7.3\sigma$ with twenty. The auto-spectra do
not fix each parameter to the precision that priors would need, but they pin down the one combination that mimics the
summed lines, to a tenth of the error on the line amplitude.

The calibrated scenario replaces a model of the continuum cross-spectrum by a model of its auto-spectrum, which
contains terms the cross-spectrum never sees: the Poisson term of the unresolved galaxies, the self-correlation of the IHL, and the Galactic and zodiacal light. This model has to meet two requirements, which we
quantify with twenty parameters over the SO footprint. First, the spectrum of the IGL Poisson term must be known to a few percent, since a tilt of $21\%$ across the band biases the summed-line amplitude by $3.0\sigma$. The forecast of
Ref.~\cite{EuclidCollab2026} leaves the shot-noise amplitude of each band free instead. Second, the diffuse Galactic
light, which adds power to the auto-spectra but only variance to the cross-spectrum, must be part of the model. A
Galactic spectrum whose slope is off by $0.1$ then biases the summed lines by $1.1\sigma$, and a slope off by $0.5$ biases them by $9\sigma$. Using only the auto-spectra above $\ell = 1000$, where the Galactic light is weak, reduces the latter to
$2.5\sigma$, at a detection of $7.1\sigma$. A continuum model of the wrong form, such as the luminosity function evaluated in the observed band, biases the lines by more than a hundred standard deviations.

In addition, the calibrated scenario treats the cross- and auto-spectra as independent
data, although both are measured from the same maps. A likelihood that keeps their correlation, treating $\kappa$ and
the channel maps as one Gaussian field, as Ref.~\cite{Cheng2024} does for multi-line intensity mapping, would be
optimal for Gaussian fields and an exact model, and would raise the detection to $9.7\sigma$. Its whole gain,
however, comes from the model's channel covariance, which it uses to cancel the sample variance of the
cross-spectrum, so that a percent of Galactic power missing from that model biases the lines by hundreds of standard
deviations. We therefore quote the independent treatment. The fit also treats the auto-spectra of the continuum as Gaussian, a poor description of the IHL
auto-spectrum, which a handful of nearby halos dominates (Sec.~\ref{sec:xkcaveats}). We have not quantified how this
affects the calibration.

\subsection{Dependence on the modeling}
\label{sec:snrrobust}

The line detection depends strongly on the line model (Table~\ref{tab:variants}). The detection of the summed lines ranges over a factor of seven to eleven, from the UniverseMachine model, whose star-formation-rate density in our halo model lies
well below the history of Ref.~\cite{HopkinsBeacom2006}, to the simple model of the simulated maps. In every model
the continuum dominates the cross-spectrum, and the lines are detected only when it is fitted. In the baseline and the fainter models this takes the all-sky survey, while the three brighter models reach $4.6$--$8.1\sigma$ in the deep fields. Even in
the UniverseMachine model, whose Ly$\alpha$ follows Ref.~\cite{Shen2025}, H$\alpha$ alone in the deep fields is
eleven times as detectable as Ly$\alpha$. \mbox{Calibrating} the star-formation rate of the baseline over all halo masses,
rather than up to the $10^{13}\,h^{-1}M_\odot$ of Ref.~\cite{Gong2017}, places more star formation in strongly biased
halos and raises the summed lines to $2.3\sigma$ in the deep fields and $9.0\sigma$ over the footprint.

\begin{figure}[tb]
  \includegraphics{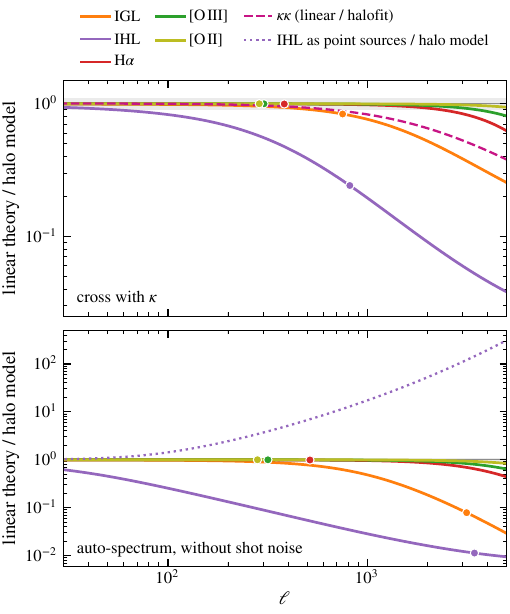}
  \caption{Ratio of the linear construction of Ref.~\cite{Shen2025} to our halo model at $2.2\um$. Top: the
  $\kappa$-cross-spectrum of the IGL, the IHL and the three lines H$\alpha$, \OIII{} and \OII{}, with the two-halo
  terms at their large-scale limit and no one-halo
  term, and the lensing auto-spectrum (dashed, linear theory over halofit). Bottom: the auto-spectra without their
  one-halo terms and shot noise (solid) and, for the IHL, with its one-halo term replaced by its white point-source
  value (dotted), the form of the simulated maps of Paper~II~\cite{PaperII}. A dot marks the multipole below which half of the ${\rm SNR}^2$ of each component accumulates in the deep fields, for the cross-spectrum with SO lensing noise (top) and for the auto-spectrum (bottom); the gray band is $\pm10\%$.}
  \label{fig:nonlinear}
\end{figure}

Linear theory, as used in the forecasts of Ref.~\cite{Shen2025}, suffices for a line on its own but not for the continuum,
nor for the lines measured against it. Our halo model has the same linear two-halo term, and adds the one-halo and Poisson terms, which Ref.~\cite{Shen2025} leaves out of the line auto-spectrum. Figure~\ref{fig:nonlinear}
therefore shows what these terms add, not the nonlinear clustering of the two-halo term, which Paper~II~\cite{PaperII} measures on its simulated maps (Sec.~\ref{sec:components}). In the Limber approximation, linear theory moves the signal-to-noise ratio of H$\alpha$ alone over the SO footprint only from $33.7$ to $33.6$, since its detectability builds up at $\ell \approx
250$--$450$, where the line traces the linear matter field. The continuum cross-spectra accumulate half of their ${\rm SNR}^2$ above $\ell \approx 800$, where much of the signal is a one-halo term, the correlation of each galaxy's
light with the mass of its own halo. Without it, linear theory lowers the signal-to-noise ratio of the IGL alone from $203$ to $169$ and that of the IHL from $146$ to $84$. For the summed lines the effect is reversed, because their fit is limited by the small-scale
fluctuations of the continuum, its one-halo and shot-noise power. Linear theory without shot noise makes the continuum
appear quieter than it is, and doubles the line detection over the footprint, from $7.3\sigma$ to $15.0\sigma$; with the shot noise kept, linear theory gives $6.2\sigma$.

These conclusions are robust to four further choices: the masking depth, the covariance between components, the diffuse Galactic light and the spectral diversity of galaxies. Masking resolved galaxies more deeply removes the line
emitters with the continuum. Masking to $26$\,AB at $1\um$ cuts the IGL cross-spectrum at $2.2\um$ to $0.07$ of its
baseline value, but the summed lines in the deep fields fall from $2.0\sigma$ to below $1\sigma$. The covariance between
components, absent from the baseline, raises the summed lines to $2.5\sigma$ in the deep fields and $8.3\sigma$ over the footprint, mostly through the coherence of different lines emitted by the same shell. This gain is an upper bound,
since every line luminosity is a fixed multiple of the halo mass. The diffuse Galactic light, added as a channel covariance of
rank one with $C_\ell \propto \ell^{-3}$~\cite{Zemcov2014} at the amplitude of the measurement of the Cosmic Infrared Background Experiment (CIBER) quoted in
Ref.~\cite{ChangLidz2026}, lowers the summed lines from $2.0\sigma$ to $1.4\sigma$ in the deep fields and from $7.4\sigma$ to $7.2\sigma$ over the footprint, and ten times this power lowers them only slightly more. Finally, real galaxies differ in spectrum, which lowers the coherence of the continuum across
channels. We model this by replacing each continuum covariance by $(1-\epsilon)\,C +
\epsilon\,C^{\rm dec}$, where $C^{\rm dec}_{ij} = (C_{ii}C_{jj})^{1/2}\mathcal{K}_{ij}$ keeps the variance of every
channel and $\mathcal{K}$ is either the identity (a white residual) or a Gaussian in $\ln\lambda$ of width $0.1$ (a
smooth residual). Over the footprint, the summed lines lose no more than about a fifth up to $\epsilon = 10^{-2}$,
while in the deep fields the smooth residual at $\epsilon = 10^{-2}$ halves them.\looseness=-1

\subsection{Which modes to remove}
\label{sec:channelmodes}

Instead of fitting templates, an analysis can remove smooth modes from each LoS, and the high-pass filter of
Sec.~\ref{sec:filter} is one such removal. In this subsection, we compare it with the two alternatives used in the
SPHEREx literature, and ask which removal keeps the most line signal for a given suppression of the continuum. For a
given set of removed directions, the statistic of Eq.~\eqref{eq:crosssnr} is already the optimal combination of
channels, so the only freedom is the choice of the removed subspace. This subspace alone sets the detection, since the quantity the filter acts on, or a taper of the band, changes the surviving fractions of Sec.~\ref{sec:filtered} but not
the signal-to-noise ratio. With a trusted model of the continuum, the optimum is to remove nothing and fit the
continuum, as the joint fit does. Removing directions makes the result robust to errors of that model. The cost of a removal depends on how much the channel directions along which the continuum can vary overlap the profiles of the lines, as measured by fitting the physical parameters of the continuum in Sec.~\ref{sec:snrshape}.

Each removal is a projection of the form of Eq.~\eqref{eq:projection}, with the columns of $Q$ spanning $n$ directions
of the 102 channels. The first basis is that of the high-pass filter, the constant and the $m$ lowest cosine and sine modes of
the band, so that $n = 2m+1$. The second is the set of \mbox{Legendre} polynomials of degree $0$ to $n-1$ in $\ln\lambda$,
the continuum fit along each LoS of Refs.~\cite{Gong2017,Cheng2020}. The third is the set of the $n$ leading
eigenvectors of the channel covariance of the continuum, averaged over multipoles, in the spirit of the
principal-component removal of Gao et al.~\cite{Gao2026}, as used in Ref.~\cite{Cagliari2025}. The first two bases remove whatever varies smoothly across
the band, including the slowly evolving cross-spectrum profiles of the lines, while the third targets the directions in
which the continuum fluctuates.\looseness=-1

\begin{figure}[tb]
  \includegraphics{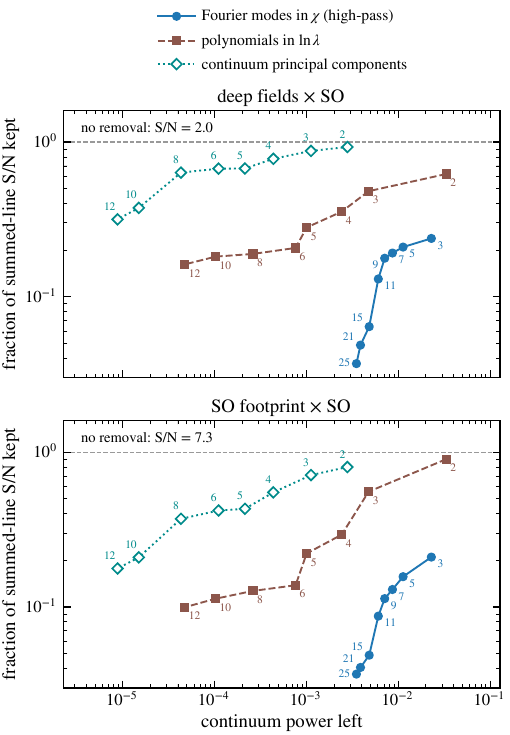}
  \caption{Fraction of the signal-to-noise ratio of the summed-line $\kappa$-cross-spectrum that survives the removal
  of $n$ smooth modes from every LoS, against the fraction of the fluctuation power of the continuum left in the data
  (the trace of its filtered channel covariance over the unfiltered one, weighted by the mode counts of the deep
  fields), in the deep fields (top) and over the SO footprint (bottom), with SO lensing noise. Each panel gives the signal-to-noise ratio without removal (Limber line spectra, as in Table~\ref{tab:variants}), the fraction's denominator. Circles: Fourier modes along the
  comoving distance (the high-pass filter of Sec.~\ref{sec:filter}). Squares: Legendre polynomials in $\ln\lambda$. Open diamonds: the $n$ leading
  principal components of the continuum's channel covariance. Each point is labeled by $n$, and the high-pass filter at $\Lambda = 0.01\Mpcinv$ is $n = 25$.}
  \label{fig:channelbases}
\end{figure}

To compare the three removals at equal suppression, in Fig.~\ref{fig:channelbases} we show the fraction of the
summed-line signal-to-noise ratio that survives each removal, against the fraction of the fluctuation power of the
continuum left in the data. The surviving fractions in the deep fields agree within a factor of two with those over the SO footprint, where the signal-to-noise ratio itself is two to five times higher. The area multiplies the
information in every channel direction alike, so that only the depth can change the fractions, and the shallower
all-sky maps lose a somewhat larger fraction once more than a few modes are removed. At every level of suppression, the high-pass filter keeps the least line signal. At $\Lambda = 0.01\Mpcinv$ it keeps a few
percent of the summed-line signal-to-noise ratio, whereas three Legendre modes keep about half of it and three
principal components most of it, with a comparable or smaller part of the continuum left in the data. The principal
components reach a given suppression with the fewest modes, because the channel covariance of the continuum is almost
two-dimensional. The directions they remove are those in which the variance of the continuum swamps the lines, and
which carry little information about them. The fraction kept by the high-pass basis, in contrast, falls steeply with $n$, since each
additional pair removes part of the slow, in-band profile of the line cross-spectra (Sec.~\ref{sec:filtered}). A search for the best removal at each level of suppression, among the directions with the
most continuum power per unit of line information, shows that the principal components are close to optimal. At the
same level of suppression, the best removal found keeps about a tenth more of the summed-line signal-to-noise
ratio, and over the SO footprint it keeps no more than the principal components once two or more are removed.

The spectral diversity of real galaxies may change this ordering on the sky. The simulated maps of
Paper~II~\cite{PaperII} cannot test this, since their continuum is as coherent across the channels as that of our
model. On these maps, the principal components keep two to three times the summed-line signal-to-noise ratio that the
high-pass filter keeps at matched or better suppression of the total IGL and the IHL. When the components are estimated
from the maps themselves, which include the lines and the noise, the fitted amplitude of the lines does not change
measurably. Reference~\cite{Roy2026a} built maps from a semi-analytic lightcone of $2\,{\rm deg}^2$, in which every
galaxy carries its own star-formation history, metallicity and dust. Removing about twenty principal components of 91
channels gave them the best balance between keeping the lines and suppressing the continuum at every multipole they
tested, $\ell = 2000$--$14\,000$. Our model needs three, since it assigns one spectrum to the unresolved galaxies at
each redshift and one to the IHL, so that three components leave about a thousandth of the fluctuation power of the
continuum (Fig.~\ref{fig:channelbases}). The two analyses also remove components for different ends.
Reference~\cite{Roy2026a} removes them until the residual maps trace the lines rather than the continuum, whereas our
joint fit describes the remaining continuum by its templates. How many components the cross-spectrum with lensing needs
on such maps, at $\ell \lesssim 1000$, has not been tested. The principal-component removal comes closest to the joint fit, the minimum-variance estimator when the
covariance is known. Three principal components cost an eighth of the signal-to-noise ratio in the deep fields and three tenths over the SO footprint, but, unlike the joint fit, the removal does not need an exact model of the \mbox{covariance} of the continuum. The high-pass filter is the poorer choice for the cross-spectrum because it removes the in-band profile of the lines along with the continuum.

In a measurement, the principal components would be estimated from the observed channel covariance, which contains
the lines and the noise. We weight each channel by its inverse noise, so the leading components are the
directions in which the continuum most exceeds the noise. Removing three of them and fitting the summed lines beside
the two continuum templates gives $1.8\sigma$ in the deep fields and $5.7\sigma$ over the SO footprint
(Table~\ref{tab:ladder}). Over the footprint this is more than the $5.2\sigma$ that three components of the
continuum's covariance alone keep in Fig.~\ref{fig:channelbases}. Estimated from 300 Gaussian draws of the data, the
components bias the summed lines by at most $0.2\sigma$ when the sky's continuum is that of the model, and the
scatter of the fitted amplitude stays within $10\%$ of the Fisher error. This route and the
calibrated scenario of Sec.~\ref{sec:snrshape} are sensitive to opposite errors. The principal components adapt to the continuum of the sky, so the luminosity function evaluated in the observed band biases the summed lines by $9.7\sigma$, rather than by more than a hundred. They do not, however, remove all the cross-spectrum of a slightly wrong continuum.
Building the IGL from the luminosity functions at the bands of Ref.~\cite{Helgason2012}, rather than from their
interpolated parameters, biases the amplitude by $11.5\sigma$, against $2.1\sigma$ in the calibrated scenario.\looseness=-1

\subsection{The auto-spectrum}
\label{sec:snrauto}

The auto-spectra are detected in a different regime. Alone in the deep fields, the IGL and the IHL reach
signal-to-noise ratios of several hundred, and the lines from $133$ for H$\alpha$ down to $0.23$ for Ly$\alpha$
(Table~\ref{tab:snrrsd}). An auto-spectrum dominated by its signal is measured at the precision of its mode
count, so the auto-spectra can calibrate the shape of the continuum (Sec.~\ref{sec:snrshape}). With every component and the white-noise level fitted jointly, the continuum loses little
and H$\alpha$ about two thirds (Table~\ref{tab:survey}). \OIII{} and H$\beta$ cannot be separated: in the four blue
bands they fall in adjacent channels at any redshift and share the same star-formation history, so that only their
blend is detected. The baseline covariance leaves out the correlation between different lines emitted by the same
shell, which fall in different channels. That correlation is the main information that the multi-line analysis of
Ref.~\cite{Cheng2024} uses. With the continuum removed beforehand, their analysis constrains the bias-weighted
intensity of H$\alpha$, \OIII{}, H$\beta$ and \OII{} at $10\sigma$ or better below $z = 2$ in the deep fields.

The line auto-spectra, however, rest on the continuum being smooth across channels. At $\ell = 1000$, $92\%$ of the
channel covariance of the continuum lies in one eigenvalue, so that a nearly diagonal line covariance can be fitted
through it at moderate cost. Any spectral diversity of the galaxies decorrelates part of the continuum across channels.
With the white residual of Sec.~\ref{sec:snrrobust}, H$\alpha$ in the joint fit falls from $47$ to $32$ at $\epsilon =
10^{-3}$ and to $15$ at $10^{-2}$, whereas a smooth residual barely moves it. We do not know how much more spectral diversity real
galaxies have than our model, so the line auto-spectra are statements about this model rather than
forecasts of the same standing as the cross-spectrum.

Unlike the cross-spectrum, the line auto-spectra gain from depth rather than area. One magnitude of additional depth
raises H$\alpha$ in the joint fit from $47$ to $155$, while the all-sky survey helps only the continuum, since the lines are
dominated by noise at its depth. As with their amplitude (Sec.~\ref{sec:filtered}), the projection costs them little.
Alone, the lines keep $72$--$80\%$ of their signal-to-noise ratio at $\Lambda = 0.01\Mpcinv$, Ly$\alpha$ the least,
close to the $[1-(2m+1)/N_{\rm ch}]^{1/2} = 0.87$ of $N_{\rm ch}$ equal, uncorrelated channels.\looseness=-1

\subsection{What the detections would measure}
\label{sec:continuumprobe}

Beyond their detection, the spectra would measure properties of the line emitters and of the two components of the
continuum.

For the lines, the cross-spectrum measures the luminosity density of the emitters weighted by their bias, near the
peak of the lensing kernel: for H$\alpha$, the star formation at $z \approx 1$--$2.5$ weighted by the bias of the halos
that host it, including faint galaxies. Surveys of resolved emitters sample only the bright part of this population,
and our line models, which differ in dust, in the normalization of the star formation and in its distribution over
halo mass, span an order of magnitude in this quantity (Sec.~\ref{sec:snrrobust}). The cross-spectrum measures it
without the shot noise of the emitters, which dominates the line auto-spectra on small scales. Over the SO footprint,
it would measure the amplitude of the summed lines to $14\%$, enough to distinguish most of these models, and that of
H$\alpha$ alone to about a third.

For the IHL, the quantity is the fraction of the light of halos that lies outside galaxies, which observations
constrain poorly (Sec.~\ref{sec:xkcaveats}). The auto-spectrum of a deep field measures it in the few most massive
nearby halos, whereas the cross-spectrum also reaches it through its two-halo term at $z \approx 0.4$: the multipoles
below $\ell \approx \XKtwohaloell$, where that term dominates, carry a fifth to a quarter of its squared
signal-to-noise ratio. Together, they would constrain the evolution of this fraction and its dependence on halo mass,
which we do not forecast here.

For the IGL, the cross-spectrum measures, channel by channel, the bias-weighted emissivity of the unresolved galaxies. It is free of the Poisson term and of any power that does not trace the matter, such as local
foregrounds and instrumental systematics, all of which enter the auto-spectrum. Comparing the correlation coefficient
$r_\ell$ of the two spectra (Fig.~\ref{fig:eblkappa}) with that of the model therefore tests what fraction of the
fluctuations traces the matter, the test of the excess fluctuations anticipated by Ref.~\cite{Feder2025}. At the precision of the all-sky survey, this test is limited by the accuracy of the template model rather than by the noise.

These applications assume a known cosmology. The same spectra also carry cosmological information, since the IGL traces
the large-scale structure. Its auto-spectrum depends on the square of its bias-weighted emissivity, its
cross-spectrum on the first power and the lensing auto-spectrum on neither, while all three scale with the
amplitude of the matter fluctuations. Analyzed together, as the clustering of galaxies and its cross-correlation
with lensing are in galaxy surveys, they would separate this amplitude from the emissivity and from the Poisson
term, which enters only the auto-spectrum. Because the spectra are measured on the same modes, a joint analysis
would also cancel part of their sample variance, as the joint Gaussian likelihood of Sec.~\ref{sec:snrshape} does for the lines, but only to the precision of the model of the auto-spectra. Foregrounds and systematics that do not
correlate between SPHEREx and the lensing map add variance to the cross-spectrum but no bias, so it tests
what the auto-spectra measure. The CIB in temperature-based lensing maps and multiplicative calibration
errors escape this test. We do not forecast these constraints here.

Cross-correlations with cosmic shear probe the same two components from the other side.
Reference~\cite{EuclidCollab2026} forecast a joint fit of the SPHEREx auto- and inter-band cross-spectra over the two
deep fields and of their cross-spectra with Euclid cosmic shear and galaxy clustering over the Euclid Deep Fields North
and South (EDF-N and EDF-S, $43\,{\rm deg}^2$), of which only EDF-N lies inside a SPHEREx deep field
(Fig.~\ref{fig:sky}). They found that the cross-spectra tighten the constraints on the IHL parameters by $10$--$30\%$, while the
IGL amplitudes, already at $1$--$2\%$ from the SPHEREx spectra alone, improve by about a tenth (their Table~A.1). In
their model as in ours, the IHL correlates more strongly than the IGL with the shear of nearby sources
(Appendix~\ref{app:shear}), so the direct gain of their cross-spectra is in the IHL, at the redshifts of the Euclid
sources. The CMB-lensing cross-spectrum is complementary in two respects. Its kernel weights the matter beyond the
peaks of the Euclid shear kernels, so it selects the IGL and reaches the IHL through its two-halo term at $z \approx
0.4$. It is also available over the whole SPHEREx sky, where the continuum is detected at high significance even with
the Planck map (Table~\ref{tab:survey}). Together, the two lensing maps would measure the IHL emissivity nearby and at
intermediate redshift, and hence its evolution.

\section{Conclusions}
\label{sec:conclusions}

We have computed the cross-spectra of the integrated galaxy light, the intrahalo light and the emission lines of the
SPHEREx channels with CMB lensing, in a common halo-model framework. We have examined how a filter along the LoS
affects them, and forecast their detectability for the SPHEREx deep fields and all-sky survey, with the lensing maps of
Planck, of the Simons Observatory and of a hypothetical lower-noise successor.

Before any filtering, the continuum dominates the cross-spectrum of every channel, and the lines contribute less than
a percent of it, because a channel collects the continuum of every redshift but each line from a single shell.
Lensing maps at different source redshifts select different continuum components, CMB lensing the integrated galaxy
light and the cosmic shear of nearby sources the intrahalo light, and so provide a tomography of the near-infrared
background.

A high-pass filter along the LoS helps separate the lines from the continuum in the auto-spectrum, but not in the
cross-spectrum with lensing. The lensing kernel is smooth, so that the cross-spectrum of a line varies across channels
as smoothly as that of the continuum, and the filter removes both alike. The lines are therefore reached by fitting the continuum out rather than by filtering it. The Limber approximation remains accurate for every filtered cross-spectrum. If the shape of the continuum across the channels is known, or calibrated by its own auto-spectra, the
summed lines reach $1.9$--$2.0\sigma$ in the deep fields with SO lensing ($0.88\sigma$ with the Planck map that covers
them) and $7.3$--$7.4\sigma$ over the SO footprint, against $1.4$--$2.3\sigma$ over the footprint if that shape is fitted from
the cross-spectrum alone. The integrated galaxy
light is detected at high significance, at $96\sigma$ already with the Planck map over the sky that its mask keeps.

The line detection depends on three elements of the model. The calibration requires a model of the continuum
auto-spectra that holds the Poisson spectrum of the unresolved galaxies to a few percent and includes the Galactic
light. The line models we considered differ by an order of magnitude in the detection. A forecast in the linear
construction of Ref.~\cite{Shen2025}, without one-halo terms and shot noise, would overstate it by up to a factor of
two. If smooth modes are removed instead of fitted, the principal components of the continuum come close to the best
removal, whereas the high-pass filter removes the line profiles along with the continuum.

These cross-spectra would also measure what the auto-spectra cannot isolate. For the integrated galaxy light, they would measure the bias-weighted emissivity of the unresolved galaxies and test how much of the near-infrared fluctuations traces the matter. For the intrahalo light, they would reach its two-halo term at $z \approx 0.4$ and measure its poorly known amplitude to a few percent. For the lines, they would measure the bias-weighted
luminosity density, for H$\alpha$ the star formation at $z \approx 1$--$2.5$. Over the SO footprint, they would measure
the amplitude of the summed lines to about $14\%$, enough to distinguish most of our line models, and that of
H$\alpha$ alone to about a third.

Several improvements would make these forecasts firmer. A measured H$\alpha$ luminosity function at $z \approx 1$--$2.5$
would fix the bright end of the line emission. The cross-correlation of SPHEREx with a spectroscopic galaxy sample at
the same redshifts would also measure the light of the fainter emitters (Paper~II~\cite{PaperII}). The covariance should include the connected trispectrum, the super-sample covariance and the local foregrounds, and the bias that the CIB induces in temperature-based lensing estimators should be evaluated. The
spectral diversity of real galaxies, which sets how well the lines separate from the continuum, is best assessed with
simulated maps in which every galaxy has its own spectrum. Reference~\cite{Roy2026a} built such maps and tested on
them the removal of the continuum from the line spectra. The cross-spectrum with lensing remains to be tested on them.

Beyond this forecast, the SPHEREx maps would allow further measurements. Analyzed together with their auto-spectra and
that of the lensing map, the cross-spectra of the continuum would constrain the amplitude of the matter fluctuations.
Lensing maps that select the intrahalo light, such as the cosmic shear of nearby sources, would pin it down. The
Euclid Deep Field North, a $20\,{\rm deg}^2$ field at the north ecliptic pole~\cite{Mellier2024}, lies inside the
SPHEREx north deep field (Fig.~\ref{fig:sky}). Euclid's slitless spectroscopy selects H$\alpha$ emitters at $z =
0.84$--$1.88$ in the wide survey and from $z = 0.41$ in its deep fields, observed also with a bluer grism~\cite{Mellier2024}. In
this field, Euclid's low-redshift shear bins would select the intrahalo light, its H$\alpha$ emitters would provide a
tracer correlated only with the emission of its own shell, and its photometry would allow fainter galaxies to be
masked.\looseness=-1

\vspace{-12pt}\penalty10000
\begin{acknowledgments}
We thank Abhishek Maniyar and Giulio Fabbian for helpful discussions regarding the prospects of measuring the cross-correlations on
the observed maps. The work of ZG and AMD is supported by the Agence Nationale de la Recherche (ANR) under grant No.~ANR-23-CPJ1-0160-01.
The authors acknowledge the use of Claude Code (Anthropic) in this work, to implement parts of the analysis code and to
prepare the manuscript text. They designed and directed the work, which builds on their earlier
pipeline, reviewed the code, verified all results, and take full responsibility for the content of this paper.
\end{acknowledgments}

\appendix
\section{Validation of the model ingredients}
\label{app:validation}

Our model combines ingredients from three published calculations, and the forecast depends on their relative
amplitudes, that of the lines against the continuum in particular. Since none of these papers provides its code, we
validate each ingredient against their published figures, which we digitize and compare in the reference's own
conventions. Each comparison uses the cosmology of its reference. In the comparison with Ref.~\cite{ChengChang2021},
we use their 96-channel design and read their printed $\Omega_m = 0.26$ as the density of cold dark matter alone, which
gives $\Omega_m = 0.309$ with the baryons. We take their Fig.~1 as unmasked and compare with their Figs.~5 and 6 under
their masks and redshift bins, but keep the halo occupation of Ref.~\cite{EuclidCollab2026} rather than theirs. We
also report the discrepancies that remain, some of which affect the forecast.\looseness=-1

The IGL luminosities follow Ref.~\cite{Helgason2012} as used by Ref.~\cite{ChengChang2021}
(Sec.~\ref{sec:components}), which publishes the IGL emissivity $d(\nu I_\nu)/dz$ with and without the intensity bias. Our IGL
emissivity agrees with theirs to within about $10\%$ over $0.75$--$5\um$ and $z = 0.25$--$4$, and the bias-weighted one
to within $\CCfiveContMax\%$ up to $z = 3$. Since their clustering model is linear, this validates the emissivity and,
although our halo occupation differs from theirs, the large-scale amplitude; neither the Poisson amplitude nor the one-halo
terms are compared. At $z = 4$ and $5$, our bias-weighted IGL
exceeds their Fig.~6 by about $15\%$, with differences of $\CCsixContMedFour\%$ and $\CCsixContMedFive\%$ in the channels free of
lines, while the emissivity itself matches their Fig.~1, so the excess lies in the bias. The different halo occupation
is a possible origin.
Below rest-frame $0.15\um$, the bluest band of Ref.~\cite{Helgason2012}, the luminosity function is unconstrained. Reference~\cite{ChengChang2021} extrapolates it below $0.15\um$ and beyond $z \approx 5$
without stating how; we recover their continuation by inverting their Fig.~1, so that the agreement there reproduces
their choice rather than testing it. This region carries $0.68\%$ of the IGL cross-spectrum at $0.76\um$ and none at
$\lamobs \ge 2.2\um$.

The clustering of the IGL and the model of the IHL follow Ref.~\cite{EuclidCollab2026}, whose kernels at $2.2\um$ we reproduce to
$\EFoneIGLat\%$ and $\EFoneIHLat\%$ rms. In comparisons with them, we evaluate the luminosity function in each observed band and mask at $20.5$\,AB, as their Table~1 and Sec.~2.1.2 imply, and use their printed IHL kernel. The IHL fraction behind their Fig.~2 is not
tabulated, so we use the best fit of their joint analysis, as in the forecast. Two of their IHL choices require an interpretation. The luminosity $L(M,0)$ of their Eq.~(4) is normalized on cluster
light in the $K$ band, whereas their text calls it a total solar luminosity. Its mass-to-light ratio at the pivot mass,
about $50\,h_{70}$ in solar units with $h_{70} = h/0.7$, is a $K$-band
value~\cite{Cooray2012,Lin2004,Lin2003}. We therefore take $L(M,0)$ in solar $K$-band units; the bolometric interpretation would raise the IHL auto-spectrum $62$-fold. Their IHL kernel is printed with $a^{1-\alpha}$. A band
luminosity, however, projects with $a^2$, as their own IGL kernel, built from observed fluxes, shows, and their
$L(M,0)$ is a band luminosity, so we use $a^{2-\alpha}$ except in reproducing their figure.

\begin{figure}
\includegraphics{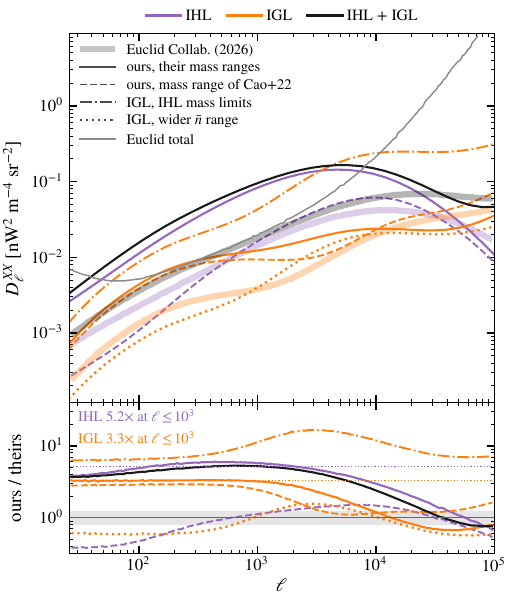}
\caption{Our IGL and IHL auto-spectra at $2.2\um$ (solid) against Fig.~2 of Ref.~\cite{EuclidCollab2026} (thick,
pale), computed in their conventions, with the ratio beneath; the horizontal dotted lines mark the mean ratios over $\ell \le 10^3$, and the gray band spans ratios of $0.8$--$1.25$. Dashed: both components on the halo-mass range of Ref.~\cite{Cao2022}. Dash-dotted and dotted: the
IGL restricted to the IHL's mass limits, and with $\bar n$ integrated over a wider mass range than the power
integrals. Black: our IGL$+$IHL, compared with the sum of their two components. Thin gray: their total, which adds
reionization, diffuse Galactic light, shot noise and instrument noise.}
\label{fig:euclidfig2}
\end{figure}

Computed in their conventions, our IGL and IHL auto-spectra at $2.2\um$ do not reproduce their Fig.~2
(Fig.~\ref{fig:euclidfig2}). Our IGL is $\EFtwoIGLavg$ times theirs, flat over $\ell = 30$--$10^3$ and close to $1/h^3$ for their $h = 0.67$. No
ingredient of the model can produce such a factor, since each enters the one- and two-halo terms at its own power. Our IHL is
$\EFtwoIHLavg$ times theirs on average over $\ell \le 10^3$, and the ratio rises from $4.2$ to $5.9$ over $\ell =
50$--$300$, which no rescaled emissivity can produce either. On the halo-mass range of the earlier model of the same first author for the China Space Station
Telescope~\cite{Cao2022}, $10^9$--$10^{13}\,h^{-1}M_\odot$, the ratio falls to $\EFtwoIHLcaoAvg$ on average, which removes
most of the amplitude offset but not its shape. We keep their stated range (Sec.~\ref{sec:components}).

The lines follow Ref.~\cite{Gong2017}, as in Ref.~\cite{ChengChang2021}. Our line power spectra agree with those of
Ref.~\cite{Gong2017} to about $10\%$ on large scales, $k \le 0.1\,h\,{\rm Mpc}^{-1}$, while on small scales, $k > 1\,h\,{\rm Mpc}^{-1}$, the difference reaches $\GFivehikhi\%$. The excess lies in the shot
noise, which weights the luminosity of the most massive halos. Every ingredient they state is matched, and the
saturation of the star-formation rate above $10^{12}\,M_\odot$ mentioned in the caption of their Fig.~2 removes the
excess at $z = 4.8$ but spoils the agreement at lower redshifts. Shot noise does not correlate with $\kappa$, so this
offset enters our forecast only through the covariance. The mean line intensities of their Fig.~3 differ from those we compute from their equations by $\GThreeZfiveMin$ to $\GThreeZloMax\%$ with redshift, by the same residual in every line.
Against the line fluxes of Fig.~1 of Ref.~\cite{ChengChang2021}, the \OIII{}$+$H$\beta$ blend agrees, but our
\OII-to-H$\alpha$ ratio is about $1.5$ times theirs. A uniform extinction of about $1.30$\,mag on every line, which
they do not mention, would remove this residual. No published value supports it, so we keep the extinctions of
Ref.~\cite{Gong2017} and treat the uniform one as a variant (Sec.~\ref{sec:snrrobust}).

\section{CMB lensing versus cosmic shear}
\label{app:shear}

\begin{figure*}[t]
\centering
\includegraphics{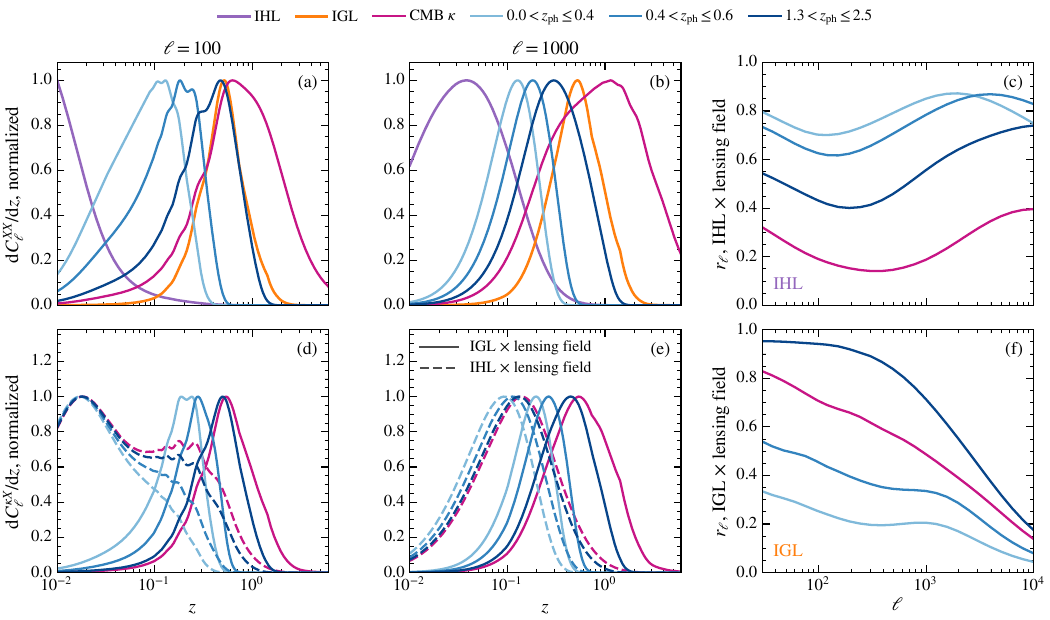}
\caption{Which continuum component a lensing field selects, at $\lamobs = 2.2\um$. (a), (b) Where each field's
fluctuations come from: $dC_\ell/dz$ of its auto-spectrum at $\ell = 100$ and $1000$, for the IHL, the IGL
(clustering term), the CMB-lensing convergence and three tomographic bins of an approximate
Euclid-like shear survey (blue, darker for higher photometric redshift). (d), (e) Where each cross-spectrum comes from:
$dC_\ell/dz$ of the IGL (solid) and IHL (dashed) cross-spectra with each lensing field, in that field's color. Every curve is
normalized to its own maximum. (c), (f) The correlation coefficient $r_\ell$ of the IHL and of the IGL with each lensing
field, signal only, in the model of Sec.~\ref{sec:unfiltered}; the IGL auto-spectrum includes its
Poisson term.}
\label{fig:shear}
\end{figure*}

In this appendix, we discuss which continuum component a lensing map selects, by comparing CMB lensing with cosmic
shear. Reference~\cite{EuclidCollab2026} forecast the cross-correlation of the same two continuum components with Euclid cosmic shear and found that the IHL correlates more strongly than the IGL, the reverse of our \mbox{ordering} with CMB lensing. They found this consistent with the intrahalo starlight tracing the gravitational \mbox{potential} more directly than their unmasked galaxies. In our \mbox{interpretation}, both orderings follow from the redshift ranges of the radial kernels: a lensing map correlates
best with the component whose emission lies at the redshifts its kernel weights. A shear kernel weights the matter \mbox{between} the observer and the sources and peaks near half their distance, so
which component it selects depends on the source redshift.

To test this interpretation, we project the same halo model against an approximate Euclid-like tomographic $n(z)$, which
is not the one of Ref.~\cite{EuclidCollab2026}. Its sources follow $n(z) \propto z^2 \exp[-(\sqrt2\,z/0.9)^{1.5}]$, of
median redshift $0.9$, with Gaussian photometric scatter $0.05(1+z)$ and no outliers, binned at $z_{\rm ph} = 0.0$--$0.4$, $0.4$--$0.6$ and
$1.3$--$2.5$. We show the result in Fig.~\ref{fig:shear}. The two low-redshift bins weight the redshifts of the IHL,
and the IGL lies beyond them (panel a). Over $\ell = 100$--$1000$, they correlate with the IHL at $r_\ell =
0.62$--$0.85$ and with the IGL at only $0.20$--$0.45$ (panels c and f). The $1.3$--$2.5$ bin weights the redshifts of
the IGL, like a shorter version of the CMB-lensing kernel, and selects the IGL more strongly than CMB lensing does. Its
correlation with the IGL falls with $\ell$ as the Poisson term of the IGL grows, and above $\ell \approx 1500$ the IHL
takes the lead in this bin too. The cross-spectra show the same overlap from the other side (panels d and e): each IGL cross-spectrum follows the kernel of its tracer, while each IHL cross-spectrum draws on the low-redshift end of it. A median redshift of $0.8$ or $1.0$, an outlier fraction of $10\%$ or a scatter of $0.03(1+z)$ moves every coefficient by at most $0.03$.

Reference~\cite{EuclidCollab2026} stated that the correlation of the IGL with the gravitational shear (their IGL $\times$ G) in their
$0.4$--$0.6$ bin is over $50\%$
lower than the IHL's. Ours is $0.73$, $0.56$ and $0.43$ of it at $\ell = 100$, $300$ and $1000$, the same direction,
though not a quantitative comparison. In our model, the ordering
reverses in their highest bin below $\ell \approx 1000$, under either power of the scale factor in the IHL kernel.

Below $\ell \approx 1000$, a SPHEREx cross-correlation with low-redshift shear is then mainly a measurement of the
intrahalo light, and one with CMB lensing or high-redshift shear a measurement of the integrated galaxy light.
Together they are a redshift tomography of the near-infrared background that the auto-spectra do not provide.
These coefficients are of the signal alone. The shape noise of a bin lowers both of its coefficients by the same
factor, so it does not change which component the bin selects. With the noise of Ref.~\cite{EuclidCollab2026}, equal
in every bin, it exceeds the lensing signal above $\ell \approx 50$ and $110$ in the two low-redshift bins, but only
above $\ell \approx 500$ in the $1.3$--$2.5$ bin, whose signal is the strongest.

\section{Projection and filtering along the LoS}
\label{app:flatsky}

In this appendix, we describe how we evaluate the flat-sky spectra, assess the accuracy of the Limber approximation,
include the redshift-space distortions of the lines, and compare conventions for the filter along the LoS. We evaluate
Eq.~\eqref{eq:flatsky} with the FFTLog method of Ref.~\cite{Gao2023} (their Sec.~2.5). The power spectrum is written
as a sum of complex power laws, $p(k) = \sum_i \alpha_i k^{\nu_i}$, and each term integrates in closed form,
\begin{gather}
  \int\!\frac{dk_\parallel}{2\pi}\, e^{ik_\parallel\delta\chi}
  \left(k_\parallel^2 + \tilde\ell^{\,2}\right)^{\nu/2}
  = \tilde\ell^{\,\nu+1} M^{(2)}_\nu\!\left(\tilde\ell\,|\delta\chi|\right), \nonumber\\
  M^{(2)}_\nu(x) = \frac{2^{(\nu+1)/2}}{\sqrt\pi\,\Gamma(-\nu/2)}\,
             x^{-(\nu+1)/2} K_{(\nu+1)/2}(x),
  \label{eq:fftlog}
\end{gather}
with $K$ the modified Bessel function of the second kind and $\tilde\ell$ as in Eq.~\eqref{eq:flatsky}. The functions
$M^{(2)}_\nu$ carry no cosmology and are tabulated once. They are finite at $\delta\chi = 0$ only for ${\rm Re}\,\nu <
-1$, so we use an FFTLog bias of $-1.6$. The two points of a pair lie at different epochs. We evaluate the power
spectrum at their mean distance and multiply its two-halo term by the ratio of the growth factors and biases at the two
epochs to their values there. Omitting this factor changes the unfiltered spectra, and the filtered cross-spectra with
$\kappa$, by less than $10^{-3}$. For top-hat shells $50$--$400$\,Mpc deep, the FFTLog evaluation agrees with a direct
numerical integration of Eq.~\eqref{eq:flatsky} to $0.05\%$ at $\ell = 60$--$2000$.\looseness=-1

\subsection{Limber accuracy}
\label{app:limber}

The Limber approximation sets $k_\parallel = 0$, which is accurate when the kernels vary slowly over $\chi/\ell$, so
that their radial Fourier content lies far below the transverse wavenumber $\ell/\chi$. The continuum and lensing
kernels are of this kind: the radial transforms of the IGL and lensing kernels fall to half power near
$5\times10^{-4}\,{\rm Mpc}^{-1}$, more than ten times below $\ell/\chi$ at $\ell = 60$. From $\ell = 50$ to $3000$, the Limber approximation therefore reproduces the auto-spectra of the IGL, the IHL and $\kappa$ to $0.3\%$, and the IGL and IHL
cross-spectra with $\kappa$ to $0.03\%$.

A single H$\alpha$ channel, by contrast, is a shell about $100$\,Mpc deep with sharp edges. Its fluctuations along the
LoS reach $k_\parallel \approx 0.03\,{\rm Mpc}^{-1}$, comparable to $\ell/\chi$ at low multipoles, so its
auto-spectrum draws on the power spectrum at $k > \ell/\chi$, where the power is lower. The Limber approximation therefore overestimates the auto-spectrum, by $42\%$ at $\ell \approx 90$ and still by $1.7\%$ at $\ell \approx 3000$.
The cross-spectrum with $\kappa$ moves by at most $0.07\%$, because the lensing kernel keeps only $k_\parallel \approx
0$. After filtering, the Limber approximation reproduces the IGL auto-spectrum to better than $0.05\%$ and the IHL
auto-spectrum to $0.3$--$6.6\%$ between $\ell = 2500$ and $60$. Its error on the line auto-spectra at $\ell = 60$ grows from
$28$--$35\%$ to $46$--$58\%$ in power, because the filter removes the smooth part of the channel covariance, where the Limber approximation holds.\looseness=-1

\subsection{Redshift-space distortions}
\label{app:rsd}

\begin{table}[t]
\setlength{\tabcolsep}{3pt}%
\fitwidth{\begin{tabular}{@{}lcc@{\hspace{20pt}}cc@{\hspace{20pt}}cc@{}}
\toprule
 & \multicolumn{2}{c}{Limber} & \multicolumn{2}{c}{flat sky} & \multicolumn{2}{c}{flat sky + RSD} \\
\cmidrule(lr){2-3}\cmidrule(lr){4-5}\cmidrule(lr){6-7}
 & alone & joint & alone & joint & alone & joint \\
\midrule
$\Lya$ & 0.25 & 0.06 & 0.20 & 0.05 & 0.23 & 0.05 \\
H$\alpha$ & 132 & 46 & 127 & 45 & 133 & 47 \\
H$\beta$ & 8.6 & 0.12 & 7.6 & 0.12 & 9.3 & 0.12 \\
\OIII{} & 59 & 1.3 & 55 & 1.4 & 60 & 1.4 \\
\OII{} & 31 & 3.1 & 29 & 2.9 & 33 & 3.1 \\
\bottomrule
\end{tabular}
}
\caption{Detectability of each line's auto-spectrum in the deep fields under three evaluations: Limber, with uncorrelated channels; the flat-sky projection in real space; and the flat-sky projection with linear redshift-space distortions (RSD), which the forecast uses. Each line is fitted alone, with only its own signal and the noise, and jointly, with every component and the noise level. The continuum terms are the same in every column.}
\label{tab:snrrsd}
\end{table}

A line is placed in a channel by its redshift, so the peculiar velocity of each emitter displaces it along the LoS.
We include the linear distortions~\cite{Kaiser1987} as in Ref.~\cite{Gao2023} (their Sec.~3.2), with the line's
two-halo integral in place of their unit bias. For channels $i$ and $j$ of one line,
\begin{multline}
  P_{ij}(k,\mu) = P^{1h}_{ij}(k) \\
  + \Big[\mathcal{I}_i \mathcal{I}_j + \big(\mathcal{I}_i f_j + \mathcal{I}_j f_i\big)\mu^2 + f_i f_j\,\mu^4\Big]
    P_{\rm lin}(k),
  \label{eq:rsdpk}
\end{multline}
with $\mu = k_\parallel/k$, $f$ the growth rate, $\mathcal{I} = (P^{2h}_{\rm line}/P_{\rm lin})^{1/2}$ the luminosity-weighted two-halo
integral, which tends to the line bias as $k \to 0$, and $P^{1h}_{ij} = (P^{1h}_i P^{1h}_j)^{1/2}$. In the cross-spectrum with $\kappa$, only the line carries the Kaiser factor. The angular terms also integrate in closed form,
\begin{align}
  \int\!\frac{dk_\parallel}{2\pi}\,e^{ik_\parallel\delta\chi}\,\mu^2 k^\nu
  &= \tilde\ell^{\,\nu+1} M^{(3)}_{\nu-2}(\tilde\ell|\delta\chi|), \nonumber\\
  \int\!\frac{dk_\parallel}{2\pi}\,e^{ik_\parallel\delta\chi}\,\mu^4 k^\nu
  &= \tilde\ell^{\,\nu+1} M^{(4)}_{\nu-4}(\tilde\ell|\delta\chi|),
  \label{eq:rsdfftlog}
\end{align}
with $M^{(3)}_\nu \equiv -d^2M^{(2)}_\nu/dx^2$ and $M^{(4)}_\nu \equiv d^4M^{(2)}_\nu/dx^4$. Since $\mu^2 k^\nu = k^\nu
- \tilde\ell^{\,2}k^{\nu-2}$, we evaluate both as combinations of $M^{(2)}$ of orders $\nu$, $\nu-2$ and $\nu-4$.

A smooth kernel, of the lensing or of the continuum, passes no radial modes on the scale $\chi/\ell$, so its product with a line shell sits at $\mu \approx 0$, and the cross-spectrum of a line with $\kappa$ moves by at most $1.1\times10^{-3}$. Thin shells also correlate with their
neighbors, which the Limber covariance sets to zero, so we evaluate every pair of a line's channels closer than
$6\chi/\ell$ at $\ell = 36$. For Gaussian redshift windows at $z = 1$ and $2.35$, the boost from the distortions agrees
with the full-sky calculation of CAMB~\cite{Lewis2000} to better than $0.1\%$ from $\ell = 40$ to $1200$.

Table~\ref{tab:snrrsd} gives the detectability of the line auto-spectra under the three evaluations. Alone, each
line is less detectable in the flat-sky projection than in the Limber approximation, by $3\%$ for H$\alpha$ to $21\%$ for Ly$\alpha$,
because the Limber approximation overestimates the auto-spectrum of a thin shell (Appendix~\ref{app:limber}). The distortions raise
the same radial modes and bring every line to within $10\%$ of its Limber value. Fitted jointly, the lines move less. The detectabilities of the $\kappa$-cross-spectra agree between the three evaluations to $0.6\%$ unfiltered and to $6\%$ at $\Lambda = 0.01\Mpcinv$, where all are below one.

\subsection{Filter conventions}
\label{app:filters}

\begin{table*}[t]
\setlength{\tabcolsep}{6pt}
\fitwidth{\begin{tabular}{lccccccccccc}
\toprule
 & \multicolumn{7}{c}{$\kappa$-cross-spectrum} & \multicolumn{4}{c}{auto-spectrum} \\
\cmidrule(lr){2-8}\cmidrule(lr){9-12}
 & \multicolumn{4}{c}{deep fields} & \multicolumn{2}{c}{SO footprint} & full sky & \multicolumn{3}{c}{deep fields} & SO footprint \\
\cmidrule(lr){2-5}\cmidrule(lr){6-7}\cmidrule(lr){8-8}\cmidrule(lr){9-11}\cmidrule(lr){12-12}
 & \multicolumn{2}{c}{SO} & S4-like & $+1$\,mag & \multicolumn{2}{c}{SO} & Planck & & & $+1$\,mag & \\
 & alone & joint & joint & joint & alone & joint & joint & alone & joint & joint & joint \\
\midrule
Ly$\alpha$ & 0.39 & 0.05 & 0.09 & 0.07 & 0.50 & 0.16 & 0.09 & 0.23 & 0.05 & 0.18 & 0.02 \\
H$\alpha$ & 16 & 0.81 & 1.2 & 1.0 & 34 & 3.0 & 2.0 & 133 & 47 & 155 & 20 \\
H$\beta$ & 4.3 & 0.04 & 0.06 & 0.06 & 5.7 & 0.10 & 0.07 & 9.3 & 0.12 & 0.49 & 0.05 \\
\OIII{} & 11 & 0.15 & 0.20 & 0.20 & 19 & 0.33 & 0.23 & 60 & 1.4 & 5.4 & 0.56 \\
\OII{} & 7.8 & 0.34 & 0.50 & 0.40 & 12 & 1.1 & 0.71 & 33 & 3.1 & 9.8 & 1.3 \\
summed lines & 26 & 2.0 & 3.0 & 2.4 & 62 & 7.4 & 4.6 & 163 & 141 & 386 & 76 \\
IGL & 30 & 23 & 39 & 24 & 203 & 173 & 96 & 435 & 373 & 482 & 2124 \\
IHL & 21 & 9.8 & 18 & 10 & 146 & 63 & 31 & 351 & 269 & 290 & 1817 \\
\bottomrule
\end{tabular}
}
\caption{Detectability of each component, unfiltered, alone and with every component fitted jointly (joint). The
$\kappa$-cross-spectrum is given in three configurations of Table~\ref{tab:surveys} (deep $\times$ SO, SO footprint
$\times$ SO and full sky $\times$ Planck) and in two variants of the deep fields with SO lensing: with the S4-like
lensing noise (S4-like), and with SPHEREx maps one magnitude deeper in every channel ($+1$\,mag). The auto-spectrum is
given in the deep fields, in their $+1$\,mag variant and over the SO footprint at the all-sky depth. The summed lines
are one template with the model's line ratios, fitted jointly beside the IGL and IHL templates. In the auto-spectrum,
the white-noise level is fitted with the components.}
\label{tab:surveyfull}
\end{table*}

The surviving fractions of Sec.~\ref{sec:filtered} depend on the quantity filtered, on the channels and on the filter,
for which we compare the projection of Eq.~\eqref{eq:projection} with an apodized high-pass filter. These choices matter only
for the continuum and the line cross-spectra: the power of the line auto-spectra changes by $3\%$ between $I_\nu$ and
$\nu I_\nu$ and by at most $11\%$ between the two filters. The quantity matters because $I_\nu =
\lambda\,\nu I_\nu/c$, so a profile smooth in one is tilted by one power of $\lambda$ in the other, across a factor of
$6.6$ in wavelength. The IHL's cross-spectrum with $\kappa$, nearly flat in $I_\nu$, keeps $0.33$--$0.71\%$ of its amplitude at
$\Lambda = 0.01\Mpcinv$ in $I_\nu$, against $9.4$--$10\%$ in $\nu I_\nu$.

The channels matter because the projection is a least-squares fit over them, and they crowd at the red end of the
band in comoving distance (Sec.~\ref{sec:filter}), so the blue end is fitted less well. In $\nu I_\nu$, the projection
keeps $6.3$--$7.0\%$ of the IGL cross-spectrum on the real channels, against $3.3$--$3.6\%$ on a uniform 256-channel axis.
Finally, the apodized high-pass filter, which tapers each profile's ends over $10\%$ of the band, passes $0.35\%$ of a
constant profile, most of what the continuum keeps under it. Its taper also removes most of what the \OIII{} and
\OII{} cross-spectra keep through the blue edge of the band. On a uniform axis of 256 channels in $\nu I_\nu$, the
apodized high-pass filter leaves $1.4$--$1.7\%$ of their amplitude, against $3.6$--$14\%$ under the projection on the same
axis. On the real channels in $I_\nu$, the projection keeps $6.0$--$20\%$ (Table~\ref{tab:survival}).\looseness=-1

\section{Comparison with Shen, Kokron and Schaan}
\label{app:shen}

In this appendix, we compare our implementation of the statistic of Ref.~\cite{Shen2025} with their published
forecasts, and examine their correction for the Limber approximation. We rebuild their forecasts from the
ingredients of their public code.\footnote{\url{https://github.com/DelonShen/LIMxCMBL}} After dividing by their ratios
for the noise-dominated and Limber approximations (their Figs.~13--14), we reproduce their SPHEREx $\Lya$ forecast to within
$3\%$ at every filter level, and their COMAP CO(1--0) forecast to within $2\%$. These reproductions use their lensing
noise, the temperature-only curves of Sec.~\ref{sec:specs}. They filter the continuous profile before binning it, so
that they must integrate the filtered covariance of the intensity maps numerically, and they set its small negative
eigenvalues to zero (their App.~D6). Since we filter the channel values, our filtered covariance is a matrix product,
positive semi-definite by construction (Sec.~\ref{sec:snrstat}).

Their Fig.~14 shows the Limber signal-to-noise ratio of the filtered cross-spectrum falling below their beyond-Limber value as the
filter strengthens, by up to $6\%$ for SPHEREx and $35\%$ for COMAP. On that basis they recommended against the Limber
approximation for foreground-filtered intensity maps (their App.~B2).

For a cross-spectrum with CMB lensing, the two evaluations should agree, filtered or not (Sec.~\ref{sec:filtered}). The survival of the cross-spectrum through lightcone evolution, the central result of
Ref.~\cite{Shen2025}, is thus fully captured by the Limber approximation, and only the size of their beyond-Limber correction is at issue. For our forecast, that correction is below $0.4\%$ for every component.

The same holds in their own two cases. Evaluated to convergence by two independent routes, the beyond-Limber signal-to-noise ratio equals the Limber value at every $\Lambda$, to $3\times10^{-4}$ for SPHEREx Ly$\alpha$ and $3\times10^{-5}$
for COMAP. Their deviation comes from the discretization of their two integrals along the LoS. They compute the filtered cross-spectrum as the difference of the unfiltered and low-passed terms, taken at the level
of the integrands (their App.~D1--D2). The unfiltered term is read from a linearly interpolated table, and the
low-passed term is integrated by the trapezoidal rule in the logarithm of the pair separation, so that the
difference retains a small fraction of the unfiltered cross-spectrum. Two discretization errors contribute: the linear interpolation of their correlation table in the logarithm of the pair separation, which only the unfiltered integral uses, and the smallest
separations, which only the unfiltered integral includes. Re-implemented, their route reproduces their Fig.~14 for
SPHEREx to within $1\%$, and agrees with the converged value once the two errors are removed. For COMAP, the two errors bring the Limber-to-beyond-Limber ratio of their route to $0.86$ at $\Lambda = 0.1\Mpcinv$, and the way their code cuts the line kernel at the
edges of the redshift range accounts for the remaining difference from their $0.65$.

\section{Detectability of every component}
\label{app:surveytable}

In this appendix, we extend Table~\ref{tab:survey} to every component (Table~\ref{tab:surveyfull}). Alone, the summed lines
reach $26\sigma$ in the deep fields and $62\sigma$ over the SO footprint, against $2.0\sigma$ and $7.4\sigma$ with the
continuum fitted (Sec.~\ref{sec:snrsurvey}). In the auto-spectrum, one summed-line template keeps $141$ of its $163\sigma$
when every component is fitted, because it removes the degeneracies between the lines (Sec.~\ref{sec:snrauto}).

\bibliography{refs}

\end{document}